\documentclass[12pt,preprint]{aastex}
\shorttitle{Candidate post T Tauri Stars}
\shortauthors{Bubar et al.}

\begin{document}

\title{KECK HIRES SPECTROSCOPY OF CANDIDATE POST T TAURI STARS}
\author{Eric J. Bubar}
\affil{Department of Physics and Astronomy, Clemson University,
    Clemson, SC 29630-0978}
\email{ebubar@clemson.edu}
\author{Jeremy R. King}
\affil{Department of Physics and Astronomy, Clemson University,
    Clemson, SC 29630-0978}
\email{jking2@ces.clemson.edu}
\author{David R. Soderblom}
\affil{Space Telescope Science Institute, 3700 San Martin Drive, Baltimore, MD 21218}
\email{soderblom@stsci.edu}
\author{Constantine P. Deliyannis}
\affil{Department of Astronomy, Indiana University, 727 East 3rd Street, Swain Hall West 319, Bloomington, IN 47405-7105}
\email{con@athena.astro.indiana.edu}
\and
\author{Ann M. Boesgaard\altaffilmark{1}}
\affil{Institute for Astronomy, 2680 Woodlawn Drive, Honolulu, HI 96822}
\email{boes@ifa.hawaii.edu}
\altaffiltext{1}{Visiting Astronomer, W.M KECK Observatory, jointly operated by the
California Institute of Technology and the University of California.}

\begin{abstract}
We use high-signal-to-noise ($\sim$150-450), high resolution ($R{\sim}45,000$) 
Keck HIRES spectroscopy of 13 candidate post T Tauri stars to derive basic physical 
parameters, lithium abundances and radial velocities.  We place our stars in the 
M$_v$-T$_{eff}$ plane for use in determining approximate ages from pre-main sequence 
isochrones, and confirm these using three relative age indicators in 
our analysis: Li abundances, chromospheric emission and the kinematic {\it U-V} plane.  Using 
the three age criteria we identify 5 stars (HIP 54529, HIP 62758, HIP 63322, HIP 74045, 
and HIP 104864) as probable post T Tauri stars with ages between 10 and 100 Myr.  
We confirm HIP 54529 as an SB2 and HIP 63322 as an SB1 star.  We also examine irregular photometric 
variability of PTTs using the {\it HIPPARCOS\/} photometry annex.  Two of our PTT stars exhibit near-IR excesses 
compared to Kurucz model flux; while recent work suggests classical T Tauri stars evince similar {\it JHK\/} excesses
presumably indicative of non-photospheric (disk) emission, our results may be illusory artifacts of the chosen {\it I\/}-band
normalization.  Near-IR excesses we see in a literature-based sample of PTTs appear to be artifacts of previous spectral
type-based $T_{\rm eff}$ values.  Indeed, comparison of the homology of their observed and model photospheric SED's suggests
that photometric temperatures are more reliable than temperatures based on spectral standards for 
the cooler temperature ranges of the stars in this sample.  We conclude that our age oriented analysis 
is a robust means to select samples of nearby, young, isolated post T Tauri stars that otherwise 
masquerade as normal field stars.

\end{abstract}


\section{INTRODUCTION}

The evolution of young stars is becoming increasingly important given the renewed 
interest in planet formation driven by the continuing discovery of exoplanets 
(e.g.\citet{2006ApJ...646..505B}).  The extremely young ($< 10^7$ yr) 
pre-main sequence precursors to stars like 
our own Sun, known as T Tauri stars, are believed to possess environments conducive to 
planetary formation.  Auspiciously, T Tauri stars have 
distinct observational characteristics that make them relatively easy to identify.  
Such characteristics include, but are not limited to, strong infrared excesses, strong 
H$\alpha$ emission, irregular variability and high surface lithium abundances (for a 
recent review of T Tauri stars refer to \citet{2003Ap.....46..506P}
and references therein).  By studying young stellar associations, T Tauri stars are fairly 
easily identifiable.  However, the period of evolution (10$^7$ to 10$^8$ yrs) between the
T Tauri and zero-age main-sequence phases remains ambiguous.  The late-type stars in this age 
range, referred to as post T Tauri stars (PTTs), are expected to have some degree of 
``intermediary characteristics'' between T Tauri and main sequence stages (\citet{1978ppeu.book..171H}).  Consequently, a 
post T Tauri star should possess, for example, relatively high measures of chromospheric 
emission and relatively high surface Li abundances compared to main sequence stars of comparable mass.
It may seem that finding these stars should be relatively simple.  By looking near regions of 
T Tauri stars, we should
find that some have evolved into their post T Tauri stage. In fact, as 
\citet{1978ppeu.book..171H} noted nearly 30 years ago, there must be many times more 
post T Tauri stars than there are T Tauri stars.  Yet studies of young stars which 
typically concentrate on regions of active star formation often 
fail to find the numbers of PTTs that are expected.  

Several explanations that could resolve this apparent disparity are discussed in 
\citet{1998ApJ...498..385S}, but here we will concentrate on their second solution: that 
PTTs exist, but are far from star-forming regions.  To summarize their findings, 
\citet{1998ApJ...498..385S} showed HD98800 to be a unique PTT system.  This 
system of four stars was determined to be a young ($\sim$10 Myr) post T Tauri 
group that is far from any region of active star formation.  The orbital properties of 
the stars mean the system could not have been violently removed from its birthplace.  
It departed by a more gentle means, suggesting that it is possible that other post T Tauri 
candidates evolved similarly and now exist in isolated environments.  

Most recently, post T Tauri search efforts have uncovered loose associations and moving groups that
appear to be of post T Tauri age (the $\beta$ Pictoris moving group, the TW Hydrae Association, 
the Tucana/Horologium Association, and the AB Doradus Moving Group)(\citet{2004ARA&A..42..685Z} and references therein).  In fact, 
HD98800 is now believed to be a member of just such a group (the TW Hydrae association).  While it is 
beginning to appear that post T Tauri stars exist in associations, the sparseness of these groups and 
their relatively large spread throughout the sky makes their identification problematic.  The salient question, then, is how 
to find these isolated PTTs in the field, away from any regions of star formation.  Without a logical starting point, 
these stars simply cannot be found other than through serendipitous means.  In this paper, though, we 
use an age-oriented definition drawing from the recommendations of \citet{2001ASPC..244....3J}.  
For our purposes, we define a post T Tauri star as a young, 
low mass, pre-main sequence star with an age in the range from 10$^7$ to 10$^8$ years, where emergence of a stellar core at 
visible wavelengths is taken as an age of zero following the work of \citet{1997MmSAI..68..807D}.  This 
age-based definition is also useful by preventing incorrect classification of a 
weak-lined T Tauri (wTTS) or naked T Tauri (nTTS), which can have features comparable 
to a PTTs, but may be significantly younger.

The objects studied here are part of our ongoing survey of solar-type stars within 60pc 
based on low-resolution (R${\sim}2000$) spectra mostly from KPNO Coude Feed.  For the 
purpose of follow-up high resolution spectroscopic programs, we flagged stars  
a) residing significantly above the {\it HIPPARCOS\/}-based main sequence and/or {\ }b) 
apparently evincing significant \ion{Ca}{2} HK emission in raw 2-d spectra inspected by 
eye at the telescope as potential PTT candidates.  

Here we present high-resolution echelle spectroscopy for 13 of these stars.  We will derive 
qualitative age information from three diagnostics (Li abundance, chromospheric emission, 
and {\it UVW} kinematics) for use in assigning post T Tauri candidacy.  We will also determine 
approximate ages from pre-main sequence isochrones and masses from pre-main sequence mass 
tracks, both taken from \citet{1997MmSAI..68..807D} (assuming solar metallicity).  For completeness, we will also search 
for indications of irregular variability, examine infrared excesses, and present equivalent 
widths for the H$\alpha$ feature: all characteristics noted by \citet{1978ppeu.book..171H} 
to be useful for identifying post T Tauri stars.  We present our interpretations of 
these three features but choose not to include them in our age-oriented analysis as they typically 
decay on shorter timescales than our chosen indicators, making them less robust.

\section{DATA AND ANALYSIS}

\subsection{\emph{Observations and Reductions}}
Spectroscopy of our objects was obtained on May 28 and 29, 2000 using the W.M. Keck I 10-m 
telescope (Table \ref{observing}), the HIRES echelle spectrograph, and a Tektronix $2048{\times}2048$ CCD 
detector.  The instrumental setup yielded a resolution of $R{\sim}$45,000 (3.2 pixel FWHM) and 
(incomplete) wavelength coverage from 4400 to 6800 {\AA}.  Standard reductions including
bias correction, flat-fielding, scattered light correction, order extraction, and 
wavelength calibrations (with rms residuals of ${\sim}0.002$ {\AA}) were carried out 
using standard routines in the {\sf echelle} package of IRAF{\footnote{IRAF is 
distributed by the National Optical Astronomy Observatories, which are operated by the 
Association of Universities for Research in Astronomy, Inc., under cooperative 
agreement with the National Science Foundation.}.   Measured per-pixel S/N values near 
the \ion{Li}{1} ${\lambda}6707$ region ranged from 150-450 and are listed in Table \ref{observing}.  
Sample spectra in the lithium region can be seen in Figure \ref{spectra}.

\subsection{\emph{Basic Physical Parameters}}
Temperatures were determined from photometric calibrations of \citet{2005ApJ...626..465R}.  
Both B-V color indices from the Tycho catalog \citep{1997hity.book.....P} and V-K$_{2mass}$ 
indices \citep{2003tmc..book.....C} were used to find effective temperatures.  Errors for 
temperatures were found from the errors in the respective color indices and in the polynomial
fits (Table 2 of \citet{2005ApJ...626..465R}).  A weighted mean 
was taken for input into Kurucz model atmospheres.  Our sample, colors, temperatures and other data are
compiled in Table \ref{basic}.

Special methods had to be utilized for one of the stars in the sample (HIP 54529) as it was found 
to be a double-lined spectroscopic binary.  Following the analysis of \citet{1986ApJ...303..724B}, 
we performed a least squares fit to the ratio of equivalent widths of two \ion{Fe}{1} absorption 
lines (W(6703)/W(6705)) versus effective temperature for the stars in our sample (Fig \ref{ewt}).  We
find the fit is given by the formula 
\begin{equation}
\frac{W(6703)}{W6705}=2.066-2.225(10^{-4})T_{eff}.
\end{equation}
We used the equivalent widths of these lines in the primary and secondary components to determine their 
respective temperatures.  Multiplicative correction factors to account for unequal continuum flux contributions
from the components (as described in \citet{1986ApJ...303..724B}) were found
to be approximately unity.  We report here on the analysis for 
only the primary component.  
This technique was not applied to HIP 63322 as it was not a double-lined star.  The double
nature of this star was revealed by asymmetry in the cross-correlation peaks of our radial velocity analysis.  

Our photometric temperatures can be compared to the spectroscopic values of \citet{2004A&A...415.1153S} 
for 6 stars in the sample (footnote `a' of Table \ref{basic}).  We find that our temperatures for warmer 
stars tend to agree nicely with their spectroscopic estimates but become systematically cooler
(by as much as 250 K at 4600 K) with declining temperatures.  We suspect the spectroscopic 
T$_{eff}$ values are afflicted by the effects of overexcitation seen in young cool 
dwarfs \citep{2006ApJ...636..432S}.  For this reason we have chosen to utilize photometric temperatures.

Use of Hipparcos parallaxes allowed us to calculate ``physical'' surface gravities using:
\begin{equation}
log\frac{g}{g_\odot}=log\frac{M}{M\odot}+4log\frac{T_{eff}}{T_{eff,\odot}}+0.4V_o+0.4B.C.+2log{\pi}+0.12
\end{equation}
where M is the mass in solar masses (determined from the pre-main sequence mass tracks of \citet{1997MmSAI..68..807D}), 
V$_o$ is the apparent magnitude and $\pi$ is the parallax in arcseconds.
The surface gravity is given in Table \ref{basic}.  The average uncertainty for the surface gravity
parameter was approximately 0.20 dex.  We note that mass differences (as discussed in Section 3.2) from using tracks of 
\citet{2000A&A...358..593S} and \citet{1998A&A...337..403B} result in surface gravities that differ from
our adopted ones by 0.01-0.17 dex when using the former and 0.01-0.18 dex when using the latter.

The microturbulence parameter was determined from calibrations of 
\citet{2004A&A...420..183A}.  They derived a relationship for determining microturbulence as 
a function of effective temperature and surface gravity.  The error in this parameter was 
determined by propagating the errors in effective temperature and log\emph{g}.  The average error
in the microturbulence was found to be 0.01 kms$^{-1}$.  In order to examine the extent to which microturbulence
would effect our lithium feature we performed lithium synthesis for microturbulent velocities spanning
1 kms$^-{1}$.  We find this change to have very little effect on the lithium feature ($\le$ 0.01 dex).  

Overall metallicities were kindly provided by R. Boone and are given in Table \ref{basic}.  
These are derived based on $\chi^2$ fitting of low-res blue spectra (R$\sim$2000) with synthetic 
spectra of varying abundance \citep{2006NewAR..50..526B}.  Internal 1-$\sigma$ level uncertainties 
are believed to be $\sim$ 0.10 dex.

\subsection{\emph{Lithium Abundances}}
The basic physical parameters for each star were utilized to linearly interpolate model atmospheres 
from the Kurucz ATLAS9 atmosphere grids.  These model atmospheres were then used in 
conjunction with the comprehensive line list from \citet{1997AJ....113.1871K} to compute 
synthetic spectra with various Li abundances using the most recent version of the spectral 
synthesis tool MOOG \citep{1973ApJ...184..839S}.  The chosen models introduce an uncertainty into the
lithium abundances of 0.03 dex, determined by comparing synthetic lithium synthesis for a standard Kurucz solar
model and one developed using our interpolated model atmospheres.  The synthetic spectra were smoothed 
appropriately by convolving them with Gaussians having FWHM values measured from clean, 
weak lines (continuum depths $\leq$0.2) in multiple orders for each star using the spectrum analysis 
tool SPECTRE \citep{1987BAAS...19.1129F}.

We positively identified lithium in 7 of the 13 stars in the sample.  We derive upper limits for 
the other 6 stars.  We attempted to determine equivalent width errors based on the 
photon noise methods of \citet{1988IAUS..132..345C}, but found 1-$\sigma$ errors $\le$1.0 m\AA, which are 
significantly smaller than the uncertainty in continuum placement.  In lieu of a more rigorous 
treatment, we choose to adopt a conservative equivalent width error estimate of 4 m\AA.  
By performing additional lithium syntheses for each of the stars, we translate this 
equivalent width uncertainty to an upper limit in lithium abundance for each star.  Sample syntheses 
are presented in Figure \ref{spectra}.  

Appropriate uncertainties in the lithium fits were derived by estimating the goodness of fit by 
means of an F-test of relative $\chi^2$ values and by adding in quadrature an approximate 
uncertainty of logN(Li)$\pm$0.08 for a temperature uncertainty of $\pm$100.  The typical 
error in the lithium abundances are found to be $\sim$0.06 dex.  The synthetic lithium abundances 
are given, with their respective errors, in Table \ref{basic}.  We also plot our lithium abundances versus effective 
temperatures for both the stars in our sample and stars in the $\sim$100 Myr Pleiades cluster 
(Figure \ref{lithium}).  The Pleiades data are taken from \citet{1993AJ....106.1059S}, \citet{1996AJ....112..186J}, 
and \citet{2000AJ....119..859K} . 

\subsection{Radial Velocities and {\it UVW} Kinematics}

Radial velocities were derived via cross-correlation analysis  using the IRAF packages 
FXCOR and RVCOR.  We used HIP 90485 as our template spectrum and adopted its 
CORAVEL radial velocity of -17.4 $\pm$ 0.2 kms$^{-1}$ \citep{2004A&A...418..989N}.  Whenever possible, 
we compared our radial velocities with precise determinations from the literature.  In all cases, our 
velocities matched within the adopted errors.  These radial velocities and errors are 
shown in Table \ref{radial}.  The quoted errors are the internal uncertainties from
fitting the cross-correlation functions.  Total radial velocity uncertainties
are larger than the cross-correlation fitting uncertainties inasmuch as $\sim$2 kms$^{-1}$
intranight telluric line shifts are observed.

The cross-correlation peaks used in the radial velocity determinations confirmed that both 
HIP 54529 and HIP 63322 were members of binary systems.  This was readily apparent in the 
double lined spectrum of HIP 54529, but HIP 63322 showed no indication of double lines.  The
asymmetry of the radial velocity cross-correlation peaks showed this star's binary nature. 

Space motions were derived with an updated version of the code used by 
\citet{1987AJ.....93..864J} which accounts for covariances.  The required 
inputs, {\it UVW} Kinematics, and uncertainties are 
given in Table \ref{radial}. 

\section{RESULTS AND DISCUSSION}
	
\subsection{\emph{Post T Tauri Status Evaluation}}

We utilize three different indicators to assess the evolutionary classification of 
our stars (summarized in Table \ref{summary}).  First is the lithium abundance.  We plot in 
Figure \ref{lithium} our derived lithium 
abundances against effective temperatures for each star, and lithium abundances in the 
Pleiades cluster from \citet{2000AJ....119..859K}; they utilized temperatures and lithium 
abundances derived from both B-V and V-I color indices which we averaged to find a single 
temperature and abundance.  Those stars with abundances that place them in the observed 
lithium abundance trend of the Pleiades are qualitatively classified as likely being young.  
Those stars which have measureable Li, yet lie below the Li trend of the Pleiades have 
inconclusive results about youth.  Finally, those stars which have no measurable lithium 
(i.e. upper limits) are stipulated to most likely be older stars, and unlikely post T Tauri candidates. 

Chromospheric activity also provides a useful estimate of youth.  In 
Figure \ref{Rhk} we plot the chromospheric activity index logR$^{\prime}$$_{HK}$ (from the recent 
chromospheric \ion{Ca}{2} H and K survey of nearby (d $\le$ 60 pc) late F through early K dwarfs of D. 
Soderblom) versus color index for each of our stars.  We separate this plot into four 
distinct activity levels, following the works of \citet{1996AJ....111..439H} and 
\citet{2003AJ....126.2048G}.  We classify those stars with logR$^{\prime}_{HK}$ greater than -4.75 
as either active or very active, and therefore, young targets.  It can be noted in 
Fig. \ref{Rhk} that many of the stars have activities resting in the inactive zone, yet 
are still near to the active part.  While it is unlikely that activity variations are entirely due
to stellar variability we note that activity can be at least partially diminished by various 
effects (i.e. Maunder Minimum phases).  With this in mind, we label these stars with inconclusive ages based on 
chromospheric emission.  Those stars which rest in the very inactive category are classified 
as unlikely to be young.

Third, we utilize {\it UVW} kinematics to discern additional qualitative age information.  We 
plotted our stars in the {\it U-V} plane (Fig. \ref{UV}) for comparison with locations of both 
`early-type' groups (young main sequence stars of spectral types B-F) and 
`late-type' groups (a mixture of young and old main sequence stars of spectral type F-M) in 
figures 8 and 10 of \citet{1999MNRAS.308..731S}.  When a star clearly resides outside of 
any structures depicted as being young they are classified as likely being older stars.  
In doing so, we identified several potential members (HIP 47007, HIP 47202, HIP 104903) of the 
alleged Wolf 630 moving group of \citet{1969PASP...81..553E}.  Youth is difficult to confirm 
from kinematics alone, therefore the {\it U-V} plane was primarily used to exclude older objects. 
While no conclusive results can be determined, the {\it UVW} kinematics did provide a useful 
criteria for confirming stars with lower lithium abundance and lower activity as being old.

We considered the above 3 criteria in interpreting evolutionary status from the H-R diagram.  
In order to describe a star as being young we require 1) a measurable lithium abundance at least
as high as that observed in the Pleiades 2) classification of 
chromospheric activity as either very active or active and 3) non-membership in ``old'' 
structures in the {\it U-V} plane.  The stars which satisfied these criteria are deemed to likely 
be young stars and mass and age estimates are determined from pre-main sequence mass tracks 
and isochrones of \citet{1997MmSAI..68..807D}.  The lithium, chromospheric emission, and kinematic 
criteria suggest that some of our objects (HIP 90004, HIP 90485, HIP 104903, HIP 47007, HIP 47202)
are post ZAMS.  Ages are not derived for these 
stars, as we have successfully eliminated them from consideration as post T Tauri 
candidates.  

\subsection{\emph{Ages}}

Ages were derived in the standard manner from the most recent pre-main sequence isochrones of 
\citet{1997MmSAI..68..807D}.  Examining the positions of our stars in the H-R diagram we 
found them to lie above or very near the main sequence.  We take an age of 100 Myr as 
indicative of membership on the zero-age main sequence, which is confirmed by the coincidence of 
single Pleiades stars from \citet{2000AJ....119..859K} with the 100 Myr isochrone.

Figure \ref{pttcolormag} contains our PTT candidates, the D'antona isochrones for ages of 
10, 20, 30, 50 and 100 Myrs,  and the mass tracks from \citet{1997MmSAI..68..807D}.  Uncertainties
in age are internal and are derived entirely from H-R diagram position and do not account for systematic 
effects in the models.  Errors from convection treatments and opacity effects are not considered.  
We also examined derived ages and masses using track from both \citet{1998A&A...337..403B} and \citet{2000A&A...358..593S} 
to determine the consistency of our derived ages.  The isochrones all give similar ages within the error bars.  The mass
tracks from \citet{1998A&A...337..403B} are shifted to lower T$_{eff}$ resulting in higher mass
estimates, particularly for the lower mass stars.  Mass differences appear to be as great as 0.07 M$\odot$
around 0.80 M$\odot$ and diminish to 0.02 M$\odot$ at a mass of 1.05 M$\odot$.  The mass estimates
from \citet{2000A&A...358..593S} tracks agree well with those from \citet{1997MmSAI..68..807D}.

\subsection{\emph{Individual Stars}}

We present results for each of the stars in the sample.  Those stars which we classify as post T Tauri
are given in bold.

\emph{HIP 47007/HD 82943}.---The HIP 47007 lithium abundance (logN(Li)=2.33) lies a factor of 5-6 below the
Pleiades distribution; the abundance is more consistent with the older ($\sim$ 600 Myr) 
Hyades distribution (Fig. 5 of \citet{1995ApJ...446..203B}).  The chromospheric emission clearly 
makes this star inactive (logR$^\prime$$_{HK}$=-4.84).  The implied older age is confirmed by the {\it U-V} 
plane, where this star appears to reside in the Wolf 630 moving group region.  Its position in
the H-R diagram indicates the star is post ZAMS.

\emph{HIP 47202/HD 83443}.---This star's Li upper limit (logN(Li)$<$0.40) and low chromospheric
emission (logR$^\prime$$_{HK}$=-4.85) suggest an old designation.  Indeed, its location in the {\it U-V} plane 
is in the middle of the Wolf 630 moving group.  While the super-solar metallicity (Fe/H = 0.20) may have led to greater
standard PMS convective lithium depletion, its evolutionary status as a post ZAMS is heavily implied by its low activity,
its kinematic plane position and its HR diagram position. 

\emph{\textbf{HIP 54529A/BD +83 319A}}.---The Pleiades-like abundance of lithium in the primary was found to 
be logN(Li)=1.21; we believe the effects of continuum dilution on this abundance to be very small (a
few percent).  This is the only star for which a chromospheric emission index was not available.  
However, we compare the level of H$\alpha$ emission with that of other stars in our sample in 
Figure \ref{halpha}.  The high level of H$\alpha$ emission is comparable to the amount of emission
seen in the most active star in the sample, leading us to label this object as very active.  We
derive an age of 25$\pm^{55}_{15}$ Myr and a mass of 0.83 $\pm$ 0.09 M$\odot$ from H-R diagram 
position.  Given the young age and no negative indication of youth from the kinematics, we 
designate this star as post T Tauri.    

\emph{HIP 59152/BD +19 2531}.---This star has no measurable lithium.  We set an upper limit of logN(Li)$<$0.10.
The moderate level of chromospheric emission (logR$^\prime$$_{HK}$=-4.52) places it along the  boundary
between the active and inactive class; therefore, age results from this criterion are inconclusive.  
Position in the {\it U-V} kinematic plane is inconclusive.  Its location places it inside of the 
Sirius branch \citep{1999MNRAS.308..731S}.  For this star, the lack of evidence of youth from activity 
and kinematics and the presence of very little lithium make this star unlikely to be a good post T Tauri candidate.  

\emph{\textbf{HIP 62758/HD111813}}.---This star has a near-Pleiades lithium abundance of \\logN(Li)=1.67; its
chromospheric emission (logR$^\prime$$_{HK}$=-4.32) places it in the active category.  These two findings suggest 
possible youth not inconsistent with its location in the {\it U-V} plane.  Using the H-R diagram 
we derive an age of 47$\pm^{30}_{10}$ Myr and a mass of 0.78 $\pm$ 0.02 M$\odot$.  Recognizing that, within 
the errors, this star is nearly on the ZAMS we classify it as a PTT or ZAMS star.

\emph{\textbf{HIP 63322/BD+39 2587}}.---We find an appreciable abundance of lithium in this star 
(logN(Li)=1.60).  Its chromospheric emission ties for the highest of any of the stars in the sample 
(logR$^\prime$$_{HK}$-4.12).  Location near the Pleiades branch \citep{2005A&A...443..609S} in the kinematic plane 
is consistent with youth.  The age (45 $\pm^{15}_{13}$Myr)
derived from the location in the H-R diagram, coupled with the significant Li and chromospheric emission,
leads us to label this star as post T Tauri.  We find a rough estimate of the mass, assuming it was a single star, of
0.77 $\pm$ 0.03 M$\odot$. We also note again that the line profiles and the 
cross-correlation peaks exhibit a notable asymmetry, indicative of this star being a spectroscopic 
binary.  

\emph{\textbf{HIP 74045/HD135363}}.---HIP 74045 is the only broad-lined star of the sample, suggesting a moderate projected rotation ($\sim$15 km/s)
consistent with youth.  The substantial lithium abundance (logN(Li)=1.96) lying in the midst of the Pleiades'
distribution is also suggestive of youth.  The chromospheric emission (logR$^\prime$$_{HK}$=-4.17) places 
this star in the very active category.  Finally, this object does not reside in any older {\it U-V} kinematic plane structures.  
The location in the H-R diagram confirms youth and a PTT
classification with an age of 36 $\pm^{14}_{6}$ Myr and a mass of 0.78 $\pm$ 0.01 M$\odot$. 

\emph{HIP 87330/HD162020}.---Our upper limit to the lithium abundance of HIP 87330 \\(logN(Li)$<$-0.30)
lies a factor of 10 below the Pleiades trend, suggesting a post ZAMS age.  However, 
the chromospheric emission index (logR$^\prime$$_{HK}$=-4.12) taken from \citet{2005A&A...443..609S}, ties for 
the highest in the sample.  The kinematics of the star yield inconclusive results.  With the lack of 
correlation between the low lithium abundance and high chromospheric emission we hesitate to make any 
definitive conclusions on the nature of this star.

\emph{HIP 90004/HD168746}.---This star has an upper limit to its lithium abundance of logN(Li)$<$0.90.  The chromospheric
emission index (logR$^\prime$$_{HK}$=-5.11) is the lowest in the sample.  While its position in the {\it U-V} plane 
is inconclusive, the low lithium upper limit, the extremely low emission index and H-R diagram position
imply this star is post ZAMS. 

\emph{HIP 90485/HD169830}.---We set an upper limit on this star of logN(Li)$<$1.5.  The chromospheric emission is also 
extremely low (logR$^\prime$$_{HK}$=-4.93).    Although the position in the {\it U-V} kinematic plane is inconclusive, 
the low lithium upper limit coupled with the low emission index clearly indicate that this star's location
above the main sequence on the H-R diagram is due to its status as a post ZAMS star.

\emph{\textbf{HIP 104864/HD202116}}.---This star shows a high lithium abundance of 
\\logN(Li)=2.50, which lies just below the Pleiades trend.  The emission index of logR$^\prime$$_{HK}$=-4.37 
places this star in the active category.  The {\it U-V} kinematics do not place this star within any
old moving group structures.  While these three indications imply youth for the star,
the position in the H-R diagram shows that, within the errors, this star could reside on the main sequence.
We derive an age of 29 $\pm^{21}_{5} Myr$ and a mass of 1.03 $\pm$ 0.02 M$\odot$.  While this age makes the
star a post T Tauri candidate, its location in the H-R diagram shows that a ZAMS classification remains plausible.

\emph{HIP 104903/HD 202206}.---This star exhibits a low Li abundance (logN(Li)=1.10), placing it well below the Pleiades
distribution.  The chromospheric emission of logR$^\prime$$_{HK}$=-4.81 places it in the inactive category, implying
a slightly older star.  Indeed, the position in the {\it U-V} plane leads to 
classification of this star as a potential member of the alleged 5 Gyr Wolf 630 moving group 
of \citet{1969PASP...81..553E}.  Its position in the H-R diagram indicates the star is post ZAMS.

\emph{HIP 114007/BD -07 5930}.---The star has an upper limit lithium abundance of logN(Li) $<$ 0.30 and an extremely 
low chromospheric emission index (logR$^\prime$$_{HK}$-4.74).  The object's position in the kinematic plane is
inconclusive.  The low lithium upper limit and the low chromospheric emission, coupled with the position 
on the H-R diagram, lead to this star's classification as ZAMS or older.   

\subsection{\emph{Irregular Variability}}

We used the {\it HIPPARCOS\/} Epoch Photometry Annex to examine photometric variability in the 5 PTT 
candidates.  This tool provided all the photometric measurements from the {\it HIPPARCOS\/}
mission for the stars in our sample.  We used these to construct histograms of the reduced
chi-squared ($\chi_{\nu}^{2}$) and the real dispersion ($\sigma_{real}$) of the V magnitudes about their 
average.  We calculate the difference in the observed and expected variances as the real variance.
  
For comparison, we performed the same analysis on a sample of the 25 best solar analog 
candidates from tables 5,6 and 7 of \citet{1996A&ARv...7..243C}.  These analogs provide 
a solid baseline of inactive stars that are presumably not subject to irregular variability.  
Additionally, we performed this analysis for 15 classical T-Tauri stars from the emission line star 
catalog of Herbig and Bell \citep{1995yCat.5073....0H} with available {\it HIPPARCOS\/} data.  The solar analogs
and T Tauri stars provide the context of a large range of anticipated variability to
fit our post T Tauri stars into.  We also include 41 post T Tauri aged stars taken from the literature 
\citep{2002AJ....124.1670M} to increase the PTT sample size and compare with our candidates.  
The $\chi_{\nu}^{2}$ and $\sigma_{real}$ can be seen in Figure \ref{bins}.  

The T Tauri stars clearly show random variability.  The majority have both 
$\chi_{\nu}^{2}$ values and real dispersions nearly an order of magnitude greater than the PTTs 
and solar analogs.

The majority of the solar analogs cluster around $\chi_{\nu}^{2}$ $\sim$ 1.  This shows 
that the analogs tend to have magnitudes close to their average, i.e. that they are much 
less variable.  Furthermore, the dispersion histogram also 
shows that the analogs stay clustered close to their average magnitudes with the typical 
dispersion $\sigma_{real}\le$ 0.01 mag.

The $\chi_{\nu}^{2}$ and $\sigma_{real}$  of our post T Tauri candidates fit nicely in the 
range exhibited in the literature sample, between $\chi_{\nu}^{2}$ values of 1 and 4, 
lending further credence to their selection as candidates.  Also, notice the dispersions 
of the post T Tauri magnitudes are both lower and less widespread than those of the TTs.  
They are not as variable as their precursors.  Examining
the overall picture, notice that both the $\chi_{\nu}^{2}$ and $\sigma_{real}$ are intermediate between the
values exhibited by T Tauri stars and solar analogs as expected.

We performed a Kolmogorov-Smirnov (K-S) test of the distributions to quantitatively explore the 
differences between the histogram distributions.  The K-S test comparing the distribution of both the
real dispersions ($\sigma_{real}$ and $\chi_{\nu}^{2}$) values for T-Tauri stars and our post T Tauri stars 
revealed that the two samples were not drawn from the same distribution.  Comparing the PTT sample
from the literature with the PTTs of this paper, we found them to be drawn from a similar 
distribution.  Finally, the K-S test for our PTTs and solar analogs revealed that the two 
cumulative samples were drawn from different distributions.  The K-S tests then solidify our PTT classications
to the extent that they confirm that PTT stars have an intermediate degree of irregular variability between
T Tauri stars and solar analogs, as anticipated.

\subsection{\emph{Infrared Excess}}

For the sake of completeness we conducted a search for any irregularities in the PTT SED's, traced by
Johnson BVI$_{C}$ photometry, 2MASS JHK$_s$ photometry and IRAS and SPITZER photometry 
when available.  Considering the proximity of these stars (within 60 parsecs of the Sun), we
did not anticipate that they would be affected by insterstellar reddening.  However, to determine the extent 
to which reddening may have an affect we created a reddening sensitive 
Johnson-band color-color diagram of (J-K) versus (V-K), following \citet{1983AJ.....88..623C}.  
The stars in our sample were clearly seen to lie along a trend
of unreddened, single stars of the Hyades cluster.  This implies that they are
not susceptible to interstellar reddening.

After establishing that reddening corrections were unnecessary, we converted the relevant magnitudes to flux densities (in Jansky) for 
comparison with Kurucz model photospheric fluxes \citep{2003IAUS..210P.A20C}, normalized at I$_{C}$.  We chose 
to normalize to the I$_{C}$ magnitudes.  Two of the 5 PTTs we identified showed significant near-infrared excesses (HIP 63322, HIP 74045).  
However, we reserve judgement on the authenticity of the observed excess in these two cases because 
using a J-band normalization yields no sign of excess in any of the stars (Fig.(\ref{kurucz_norm})).  Indeed,
in contrast to the results of \citet{2005ApJ...635..422C} on JHK excesses in CTTs, we find that a J-band
normalization slightly improves the fit to the SED at other wavelengths.

In order to examine the likelihood that a PTTs would exhibit a near-IR excess, we performed the same analysis on 
a literature sample of 16 post T Tauri stars \citep{2002AJ....124.1670M}, who utilize spectral-type based
T{\rm eff\/} values.  The sample 
analyzed showed that excess was present in approximately 50 percent of the stars analyzed.  To further examine our 
methodology, we analyzed a sample of presumably unremarkable solar analogs to confirm that no 
spurious effects were present.  None of the solar analogs analyzed exhibited any form of 
infrared excess.  In Figures \ref{blackbody1} and
\ref{blackbody2} we present SEDs for our stars as well as a sample of solar analogs and 
literature post T Tauri stars.  

The observed SEDs of many objects in the literature sample
of post T Tauri stars seemed to match the morphology of our post T Tauri candidates of lower T{\rm eff\/} thus
we calculated photometric temperatures for each of the literature stars and performed a Kurucz
model flux fit using these photometric temperatures.  The model fluxes characterized by photometric temperatures
fit the observed SEDs better than those characterized by the literature-based effective temperature values (Fig. \ref{kurucz_MAMAMIA})
which are 250-1000 K higher.
This indicates that photometric temperatures may be more reliable than those determined from spectral type calibrations 
for the cooler PMS stars or the log(g)-based decrements used by \citet{2002AJ....124.1670M} are too small.

\subsection{\emph{H$\alpha$ Equivalent Widths}}

H$\alpha$ emission provides a strong indication of youth in a star.  However,
if we consider H$\alpha$ emission relative to our other indicators of youth, it has the 
smallest decay time.  So, while high levels of emission imply youth, low levels of 
emission (or high levels of absorption) do not necessarily discredit youth.  For completeness, measurements 
of the equivalent widths of the H$\alpha$ feature for each of the stars in the sample are 
presented in Table \ref{eqwidth} since these line strengths are often used in classifying CTTs.

We note that the binary star HIP 63322 exhibits a P-Cygni profile in the H$\alpha$ region, a 
feature that is common in many classical T Tauri stars; however, the EW is too low to suggest such a 
classification.  The blueshifted absorption feature 
can be attributed to a strong stellar wind, however, a small redshifted absorption feature 
also appears to be present.  Data with the H$\alpha$ line centered away from the edge of the CCD are needed
to examine this feature and 
determine if it is ``real''.  If it represents an actual absorption, this could be 
indicative of infall onto one of the members of the system, making this a particularly 
interesting example of a post T Tauri system.    

\section{SUMMARY}
We have utilized an age-oriented analysis to identify 5 isolated post T Tauri candidates (HIP 54529, 
HIP 62758, HIP 63322, HIP 74045, HIP 104864) and analyzed their irregular variability and SEDs.  The irregular 
variability of our candidates, and PTTs in general, appear to be intermediate in nature to T Tauri and solar analog 
variability.  Two of the 5 candidates (HIP 63322 and HIP 74045) exhibit near-IR excesses when normalized to the
I$_{C}$, although this same excess is non-existent with a J-band normalization.  Subsequent study
must be undertaken to determine the nature and validity of these excesses.  In our SED analysis, we also find that model 
fluxes based on photometric temperatures appear to match observed SEDs better than model fluxes using temperatures based
on spectral standards.  Also of note is the binarity of 2 of our
post T Tauri stars: HIP 54529 and HIP 63322 are found to be spectroscopic binaries.

Our combination of H-R diagram positions and various qualitative indicators of youth (including lithium abundances,
chromospheric emission and kinematics) appears to be a robust means to select samples of nearby, young, isolated post T
Tauri stars that would otherwise masquerade as normal field stars.  The method that we have developed will be
applied to larger samples of stars to further enhance the population of known post T Tauri stars.

\acknowledgments
We thank Roggie Boone for use of his metallicity determinations.  We also wish to thank the anonymous
referee whose comments helped to improve and clarify the paper.  This 
research has made use of the NASA/ IPAC Infrared Science Archive, which is operated by 
the Jet Propulsion Laboratory, California Institute of Technology, under contract with 
the National Aeronautics and Space Administration.  This publication also makes use of data 
products from the Two Micron All Sky Survey, which is a joint project of the University 
of Massachusetts and the Infrared Processing and Analysis Center/California Institute of 
Technology, funded by the National Aeronautics and Space Administration and the National 
Science Foundation.  JRK and EJB gratefully acknowledge support for this work from NSF grants AST-0086576 and 
AST-0239518.  EJB would also like to acknowledge support from the South Carolina Space Grant Consortium.

\clearpage

\begin{deluxetable}{l c c c c c c }
\tablewidth{0 pt}
\tablecaption{Log of Observations, 2000 May 28-29 UT}
\startdata
\hline
\hline
HIP     &  BD/HD         & V   & B-V	& MJD \tablenotemark{a} & Exposure Time   & S/N \\
        & 	         &     &	& 51692+	        & (sec) 	        &     \\ 
\hline
47007   &   HD 82943	 &6.53 & 0.625 & 0.263922	       & 120	    & 464 \\
47202   &   HD 83443	 &8.23 & 0.814 & 0.267599	       & 300	    & 266 \\
54529A  &   BD +83 319A  &9.93 & 0.949 & 0.276696	       & 300	    & 159 \\
59152   &   BD +19 2531  &9.19 & 0.884 & 0.282785	       & 300	    & 267 \\
62758   &   HD 111813    &9.09 & 0.907 & 0.302474	       & 180	    & 222 \\
63322   &   BD +39 2587  &9.27 & 0.855 & 0.291452	       & 180	    & 193 \\
74045   &   HD 135363    &8.72 & 0.949 & 0.369162	       & 120	    & 198 \\
87330   &   HD 162020    &9.18 & 0.968 & 0.446854	       & 300	    & 247 \\
90004   &   HD 168746    &7.95 & 0.713 & 0.452436	       & 180	    & 320 \\
90485   &   HD 169830    &5.91 & 0.518 & 0.456642	       &  60	    & 447 \\
104864  &   HD 202116    &8.39 & 0.614 & 0.610233	       & 120	    & 210 \\
104903  &   HD 202206    &8.08 & 0.714 & 0.613773	       & 300	    & 386 \\
114007  &   BD -07 5930  &9.55 & 1.069 & 0.619408	       & 180	    & 165 \\
\enddata

\tablenotetext{a}{MJD=JD-2400000.5}
\label{observing}
\end{deluxetable}

\begin{deluxetable}{l c c c c c c c c c c}
\rotate
\tablewidth{0 pt}
\tablecaption{Stellar Parameters}
\startdata
\hline
\hline
HIP                      & BD/HD       & B-V    & V-K$_{2mass}$ & M$_{V}$ & T$_{eff}$      & log(g)          & $\xi$      &logR$\prime$$_{HK}$       & logN(Li)        & [M/H]       \\
                         &             &        &               &         &                &                 & kms$^{-1}$ &                          &                 &    \\ 
\hline
47007 \tablenotemark{a}	 & HD 82943    &  0.625 & 1.425         & 4.337   & 5851 $\pm$ 49  & 4.21 $\pm$ 0.09 &  1.05      & -4.837		     & 2.33 $\pm$ 0.08 & 0.0   \\
47202 \tablenotemark{a}  & HD 83443    &  0.814 & 1.786         & 5.036   & 5313 $\pm$ 50  & 4.24 $\pm$ 0.16 &  1.26      & -4.850 \tablenotemark{b} & $\le$ 0.40      & 0.20      \\
54529A \tablenotemark{c} & BD +83 319A &  0.949 & 2.628         & 6.164   & 4668 $\pm$ 162 & 4.36 $\pm$ 0.55 &  0.98      & \nodata                  & 1.10 $\pm$ 0.06 & -0.14  \\
59152 			 & BD +19 2531 &  0.884 & 2.162         & 6.203   & 4947 $\pm$ 50  & 4.44 $\pm$ 0.19 &  1.08      & -4.523		     & $\le$ 0.10      & 0.00	 \\
62758			 & HD 111813   &  0.907 & 2.096         & 6.210   & 4990 $\pm$ 49  & 4.45 $\pm$ 0.18 &  1.10      & -4.320		     & 1.67 $\pm$ 0.06 & -0.10 \\
63322  \tablenotemark{d} & BD +39 2587 &  0.855 & 2.389         & 6.365   & 4786 $\pm$ 49  & 4.45 $\pm$ 0.22 &  1.01      & -4.116		     & 1.60 $\pm$ 0.03 & -0.20 \\
74045			 & HD 135363   &  0.949 & 2.529         & 6.375   & 4680 $\pm$ 50  & 4.41 $\pm$ 0.10 &  0.98      & -4.169		     & 1.96 $\pm$ 0.05 & -0.10 \\
87330 \tablenotemark{a}	 & HD 162020   &  0.968 & 2.641         & 6.705   & 4577 $\pm$ 52  & 4.47 $\pm$ 0.20 &  0.94      & -4.120 \tablenotemark{b} & $\le$ -0.30     & 0.00	 \\
90004 \tablenotemark{a}	 & HD 168746   &  0.713 & 1.697         & 4.777   & 5477 $\pm$ 52  & 4.04 $\pm$ 0.17 &  1.40      & -5.107		     & $\le$ 0.90      & -0.17	 \\
90485 \tablenotemark{a}	 & HD 169830   &  0.518 & 1.222         & 3.109   & 6207 $\pm$ 54  & 3.95 $\pm$ 0.12 &  1.81      & -4.925		     & $\le$ 1.50      & 0.00	 \\
104864			 & HD 202116   &  0.614 & 1.535         & 4.909   & 5735 $\pm$ 53  & 4.37 $\pm$ 0.17 &  1.39      & -4.368		     & 2.50 $\pm$ 0.04 & -0.05 \\
104903 \tablenotemark{a} & HD 202206   &  0.714 & 1.595         & 4.750   & 5581 $\pm$ 51  & 4.18 $\pm$ 0.18 &  1.39      & -4.808		     & 1.10 $\pm$ 0.11 & 0.13  \\
114007  		 & BD -07 5930 &  1.069 & 2.288         & 6.486   & 4799 $\pm$ 54  & 4.43 $\pm$ 0.23 &  1.03      & -4.735		     & $\le$ 0.30      & -0.05	 \\
\enddata

\tablenotetext{a}{ These stars are all known hosts of extra-solar planets.  They also have spectroscopic temperatures 
available in \citet{2004A&A...415.1153S} and \citet{2004A&A...414..601I}.}  
\tablenotetext{b}{ Chromospheric activity indices for these stars were taken from \citet{2005A&A...443..609S}.  All
others were taken from the recent chromospheric \ion{Ca}{2} H and K survey of nearby (d$\le$60 pc) late F and 
early K dwarfs of D. Soderblom.}
\tablenotetext{c}{ This star is a double lined spectroscopic binary (SB2).  We label it with an `A', as our analysis 
is for the primary component.  The error on the temperature was determined by taking the RMS deviation of the equivalent width
fit of Fig. \ref{ewt}.}
\tablenotetext{d}{This star is a spectroscopic binary (SB1) as revealed by asymmetry in cross-correlation peaks.} 
\label{basic}
\end{deluxetable}

\begin{deluxetable}{l c c c c c c c }
\rotate
\tablewidth{0 pt}
\tablecaption{Kinematic Infomation}
\startdata
\hline
\hline
HIP & $\pi$ & PM RA   & PM DEC  & Radial Velocity  & U          & V          & W          \\
    & mas   & mas/yr  &  mas/yr & kms$^{-1}$       & kms$^{-1}$ & kms$^{-1}$ & kms$^{-1}$ \\								
\hline
 47007 & 36.42 $\pm$ 0.84 &  2.38   $\pm$ 0.89 & -174.05 $\pm$ 0.53 &  8.53  $\pm$ 0.44	 &  10.05 $\pm$ 0.36 & -20.38 $\pm$ 0.47 & -8.53  $\pm$ 0.37\\
 47202 & 22.97 $\pm$ 0.90 &  22.35  $\pm$ 0.72 & -120.76 $\pm$ 0.69 &  29.34 $\pm$ 0.63	 &  19.94 $\pm$ 0.80 & -31.26 $\pm$ 0.63 & -11.82 $\pm$ 0.62\\
54529A & 17.65 $\pm$ 2.66 & -69.59  $\pm$ 2.80 & -34.35  $\pm$ 2.34 & -2.04  $\pm$ 10.00 \tablenotemark{a}& -15.96 $\pm$ 5.62 & -13.47 $\pm$ 6.98 & -1.32  $\pm$ 5.52\\ 
 59152 & 25.27 $\pm$ 1.40 &  120.06 $\pm$ 1.54 & -73.58	 $\pm$ 0.72 & -5.11  $\pm$ 0.55	 &  26.69 $\pm$ 1.48 & -0.47  $\pm$ 0.23 & -3.03  $\pm$ 0.55\\ 
 62758 & 26.54 $\pm$ 1.45 & -142.59 $\pm$ 1.24 & -37.49	 $\pm$ 1.19 & -4.06  $\pm$ 0.59	 & -17.79 $\pm$ 0.99 & -19.38 $\pm$ 1.09 & -4.04  $\pm$ 0.59\\
 63322 & 26.24 $\pm$ 1.75 & -126.76 $\pm$ 1.33 & -42.02	 $\pm$ 1.22 & -12.31 $\pm$ 0.85	 \tablenotemark{b}& -13.84 $\pm$ 1.02 & -19.38 $\pm$ 1.29 & -9.97  $\pm$ 0.85\\
 74045 & 33.97 $\pm$ 0.69 & -131.87 $\pm$ 0.62 & 169.32	 $\pm$ 0.79 & -7.52  $\pm$ 1.77	 & -26.21 $\pm$ 0.80 & -13.05 $\pm$ 1.29 & -9.68  $\pm$ 1.10\\
 87330 & 31.99 $\pm$ 1.48 &  20.99  $\pm$ 2.35 & -25.20	 $\pm$ 1.27 & -27.52 $\pm$ 0.59	 & -27.60 $\pm$ 0.58 &  2.55  $\pm$ 0.27 & -1.49  $\pm$ 0.38\\
 90004 & 23.19 $\pm$ 0.96 & -22.13  $\pm$ 0.90 & -69.23	 $\pm$ 0.66 & -25.51 $\pm$ 0.39	 & -19.15 $\pm$ 0.42 & -22.04 $\pm$ 0.60 & -3.11  $\pm$ 0.21\\
 90485 & 27.53 $\pm$ 0.91 & -0.84   $\pm$ 1.23 &  15.16	 $\pm$ 0.72 & -17.40 $\pm$ 0.20	 & -16.94 $\pm$ 0.20 &  1.18  $\pm$ 0.16 &  3.81  $\pm$ 0.20\\
104864 & 20.13 $\pm$ 1.16 &  101.39 $\pm$ 1.20 & -10.22	 $\pm$ 0.58 & -2.87  $\pm$ 0.34	 & -17.77 $\pm$ 0.97 & -3.31  $\pm$ 0.22 & -16.01 $\pm$ 1.07\\
104903 & 21.58 $\pm$ 1.14 & -38.23  $\pm$ 1.36 & -119.77 $\pm$ 0.49 &  13.78 $\pm$ 0.44	 &  22.11 $\pm$ 0.76 & -19.53 $\pm$ 1.31 & -9.36  $\pm$ 0.36\\
114007 & 24.39 $\pm$ 1.75 & -172.51 $\pm$ 2.01 & -201.31 $\pm$ 1.46 &  27.85 $\pm$ 0.62	 &  52.83 $\pm$ 3.39 & -7.24  $\pm$ 1.57 & -24.44 $\pm$ 0.56\\

\enddata
\tablenotetext{a}{The radial velocity is a crude estimate of the systemic velocity of the binary system.  It was taken by
averaging the velocities of the two cross-correlation peaks in the FXCOR package of IRAF.}
\tablenotetext{b}{This star is an SB1 as indicated by an asymmetric cross-correlation peak in the
radial velocity analysis.  The velocity for this system should be taken as an estimate.}
\label{radial}

\end{deluxetable}

\begin{deluxetable}{l c}
\tablewidth{0 pt}
\tablecaption{H$\alpha$ Equivalent Widths}
\startdata
\hline
\hline
 HIP    & H$\alpha$ EW \\
        &  $\AA$ \\
\hline
 47007  &  1.1368 \\
 47202  &  1.0342 \\
 54529 \tablenotemark{a} & -0.1703 \\
 59152  &  0.9546 \\
 62758 \tablenotemark{a} &  0.8549 \\
 63322  &  0.2521 \\
 74045 \tablenotemark{a} & -0.2115 \\
 87330 \tablenotemark{a} &  0.6016 \\
 90004  &  1.0409 \\
 90485  &  1.1310 \\
 104864 &  0.9736 \\
 104903 &  1.0233 \\
 114007 &  0.8626 \\
\enddata
\tablenotetext{a}{ The H$\alpha$ feature for these stars are plotted in Fig. \ref{halpha}.}
\label{eqwidth}
\end{deluxetable}

\begin{deluxetable}{l c c c c c c c}
\tablewidth{0 pt}
\tablecaption{Summary of Age-Oriented Analysis Results}
\startdata
\hline
\hline
 HIP    & logN(Li)    & logR$^\prime$$_{hk}$ & Kinematics & H-R Diagram & Status  & Ages $^a$             & $\delta(Age)$ $^b$\\
        &                    &                &            &             & 		  & Myr               &               \\ 
\hline
 47007  & yes                & no             & no         & no          & Post ZAMS 	  &                    & \\
 47202  & ?                  & no             & no         & no          & Post ZAMS      &                    & \\ 
 54529A & yes                & yes $^b$       & ?          & yes         & PTTs           & 25$\pm^{55}_{15}$  & -3.16 \\ 
 59152  & no                 & yes            & ?          & ?           & ZAMS or Older  &                    & \\
 62758  & yes                & yes            & ?          & ?           & PTTs or ZAMS   & 47$\pm^{30}_{10}$  & \\
 63322  & yes                & yes            & ?          & yes         & PTTs 	  & 45$\pm^{15}_{13}$  & -6.93\\
 74045  & yes                & yes            & ?          & yes         & PTTs 	  & 36$\pm^{14}_{6}$   & -3.38\\
 87330  & no                 & yes            & ?          & yes         & $\sim$ZAMS     &                    & \\
 90004  & no                 & no             & ?          & no          & Post ZAMS      &                    & \\
 90485  & no                 & no             & ?          & no          & Post ZAMS      &                    & \\
 104864 & yes                & yes            & ?          & ?           & PTTs or ZAMS   & 29$\pm^{21}_{5}$   & \\
 104903 & ?                  & no             & no         & no          & Post ZAMS      &                    & \\
 114007 & no                 & no             & no         & ?           & ZAMS or Older  &                    & \\
\enddata
\tablenotetext{a}{ Ages are derived based on the assumption of solar metallicity.}
\tablenotetext{b}{The quantity $\delta(Age)$ gives the difference between ages assuming solar metallicity and ages
which account for the sub-solar metallicites.  We utilized the online tools of 
\citet{2000A&A...358..593S} to interpolate between solar (Z=0.02) and subsolar (Z=0.01) stellar ages to obtain
our metallicity sensitive ages.  The negative sign indicates that metallicity sensitive ages are younger.}
\tablenotetext{c}{ We compare the levels of H-alpha emission of this star with several
other stars in the sample in Figure \ref{halpha}.  This star is seen to have emission levels
similar to those of one of the more chromospherically active stars in the sample, thus
we label it as``active''.}

\label{summary}  
\end{deluxetable}

\clearpage

\begin{figure}
\plotone{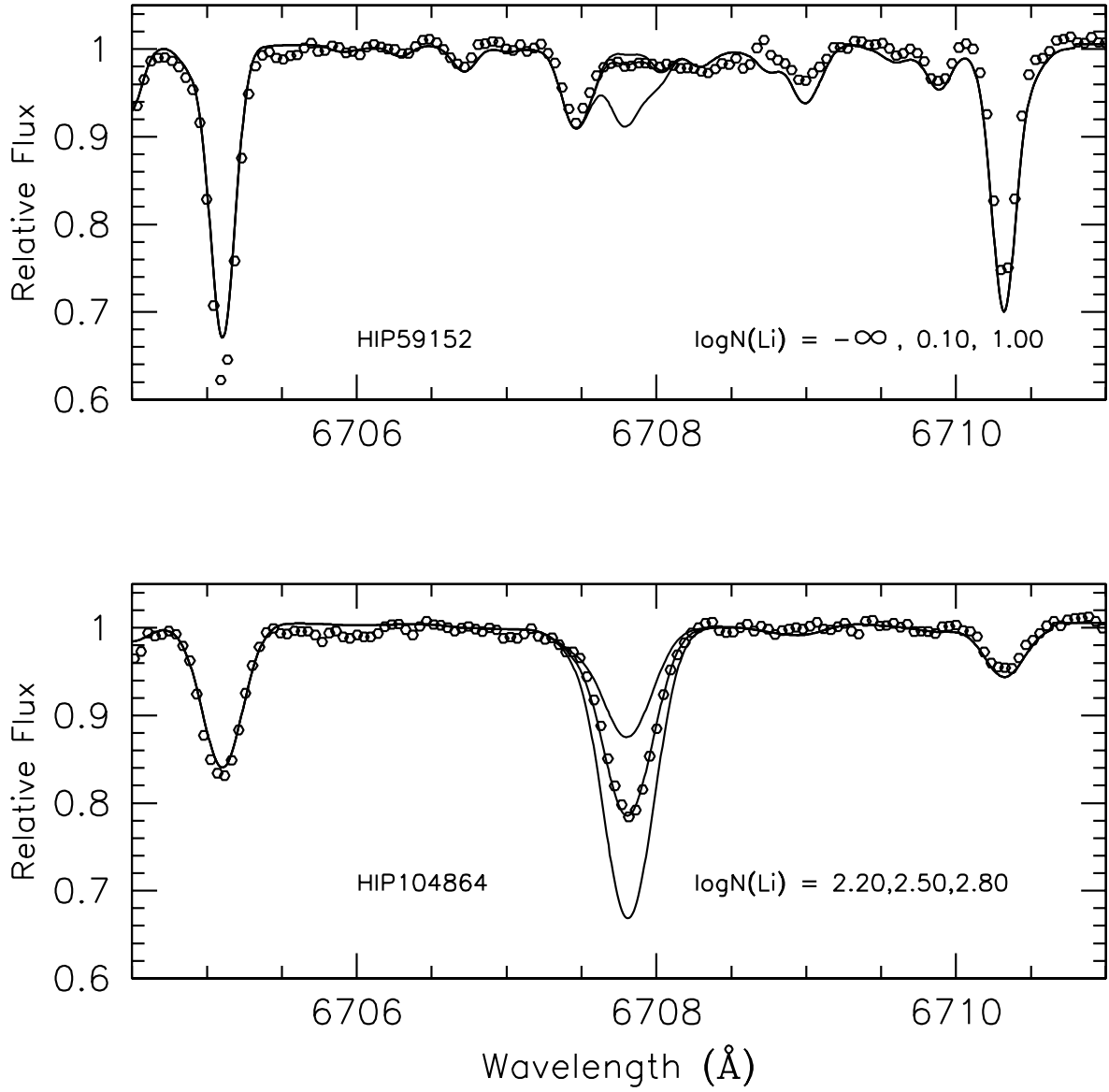}
\caption{Sample Keck HIRES spectra in the \ion{Li}{1}${\lambda}6707$ region.  The observed spectra
are plotted as open circles and varying synthesis of lithium abundances are plotted as solid lines.
HIP 59152 (top) has an upper limit (logN(Li)$\le$0.10) on its lithium abundance and HIP 104864 (bottom) is 
a clear lithium abundance determination (logN(Li)=2.50).  The varying lithium synthesis in the positive detection 
each differ by a factor of 2.}\label{spectra}
\end{figure}

\begin{figure}
\plotone{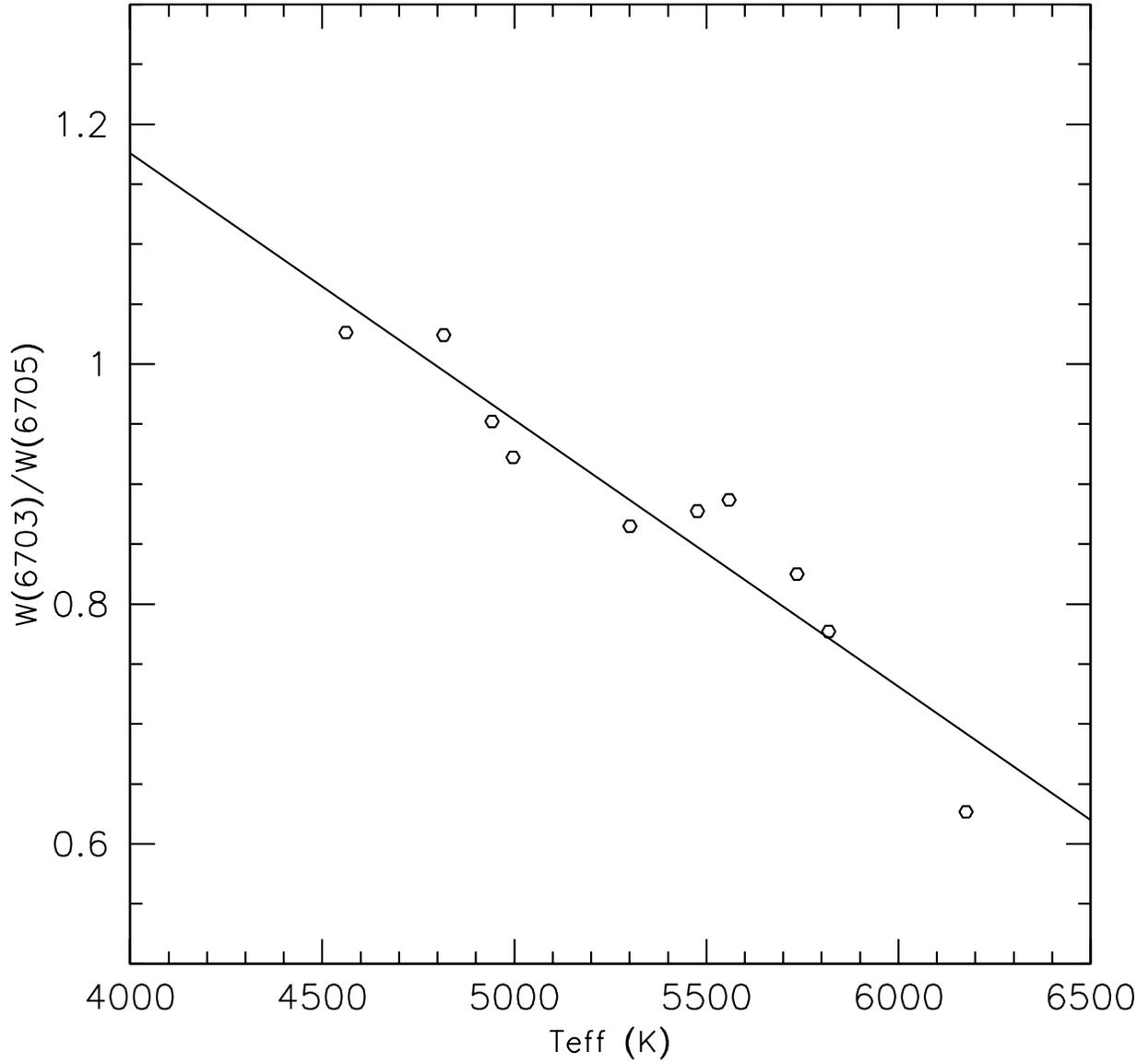}
\caption{Equivalent width ratios of two Fe I lines ($\lambda$=6703 and 6705) versus photometric 
temperatures for the stars in our sample.  The line is a least squares fit to the data which
was used for estimating temperatures in the primary and secondary components of the double-lined 
spectroscopic binary HIP 54529.  This follows the method outlined by \citet{1986ApJ...303..724B}.}\label{ewt}
\end{figure}

\begin{figure}
\plotone{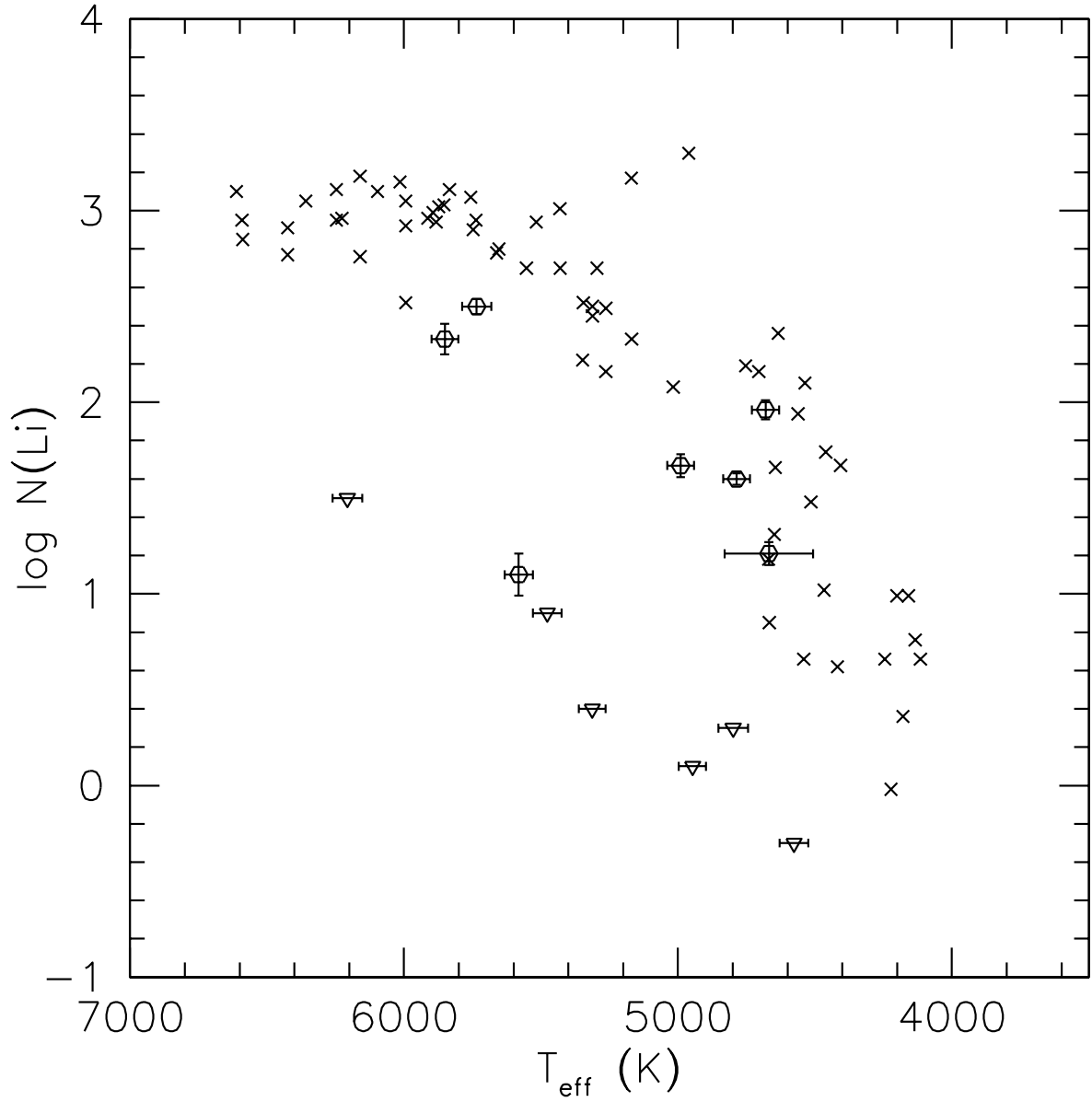}
\caption{Lithium abundances compared to those of the Pleiades star cluster.  The Pleiades stars are
plotted as x's.  Positive identifications of lithium are shown as open hexagons.  Upper limits on
lithium are plotted as triangles.}\label{lithium}
\end{figure}

\begin{figure}
\plotone{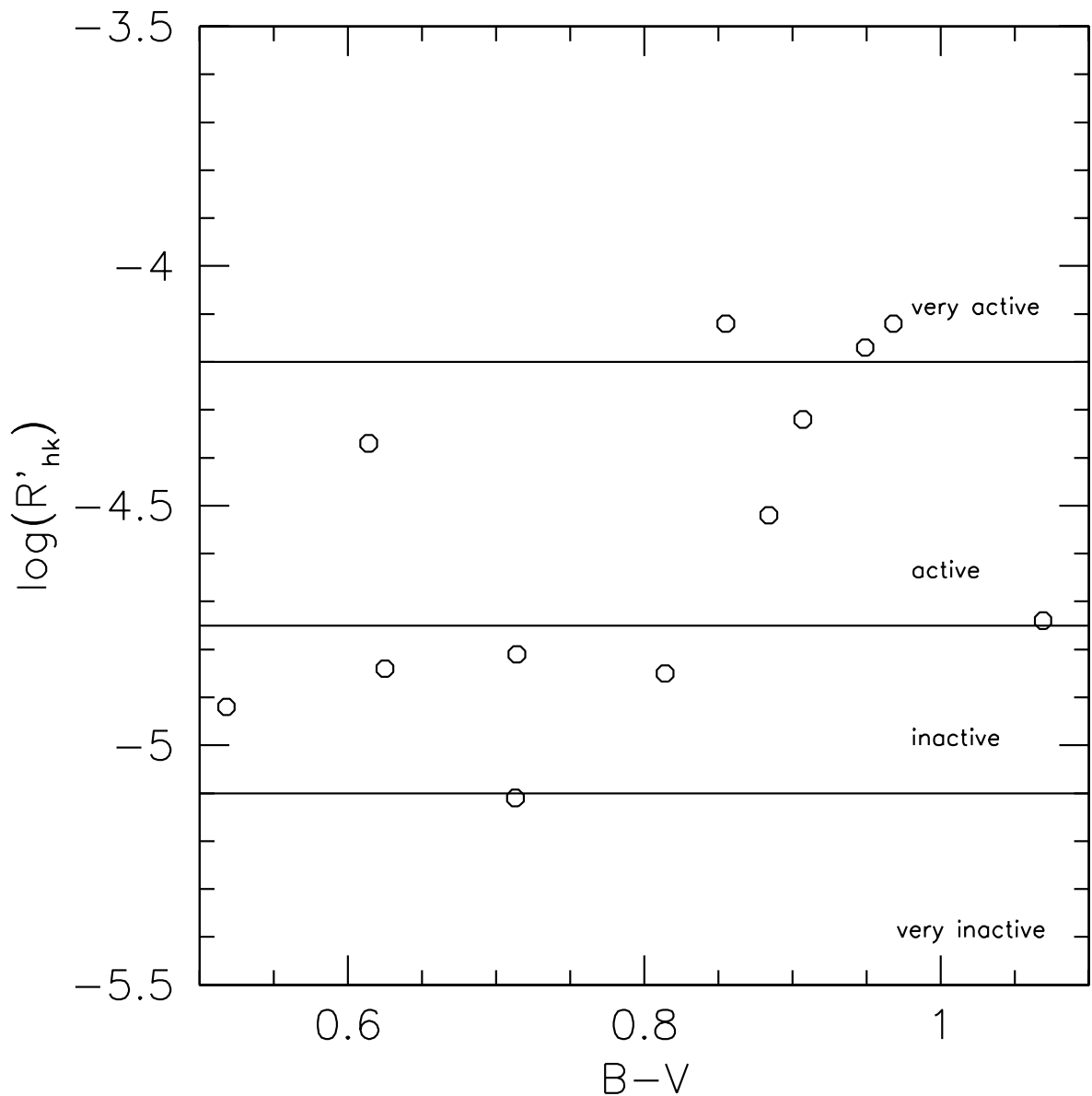}
\caption{Levels of chromospheric emission (logR$^\prime_{HK}$) given as a function of B-V color index.  The activities
are separated into four classes of activity following \citet{1996AJ....111..439H}.  Stars which are very inactive
are likely not the young pre-main sequence stars being sought.  Those stars which are either very active or active
are likely younger stars making them suitable PTTs candidates.  Due to fluctuations in activity cycles 
(i.e. Maunder minimum-like phases) we cannot rule out inactive stars as old.  They could potentially be exhibiting 
uncharacteristically low levels of chromospheric emission.}\label{Rhk}
\end{figure}

\begin{figure}
\plotone{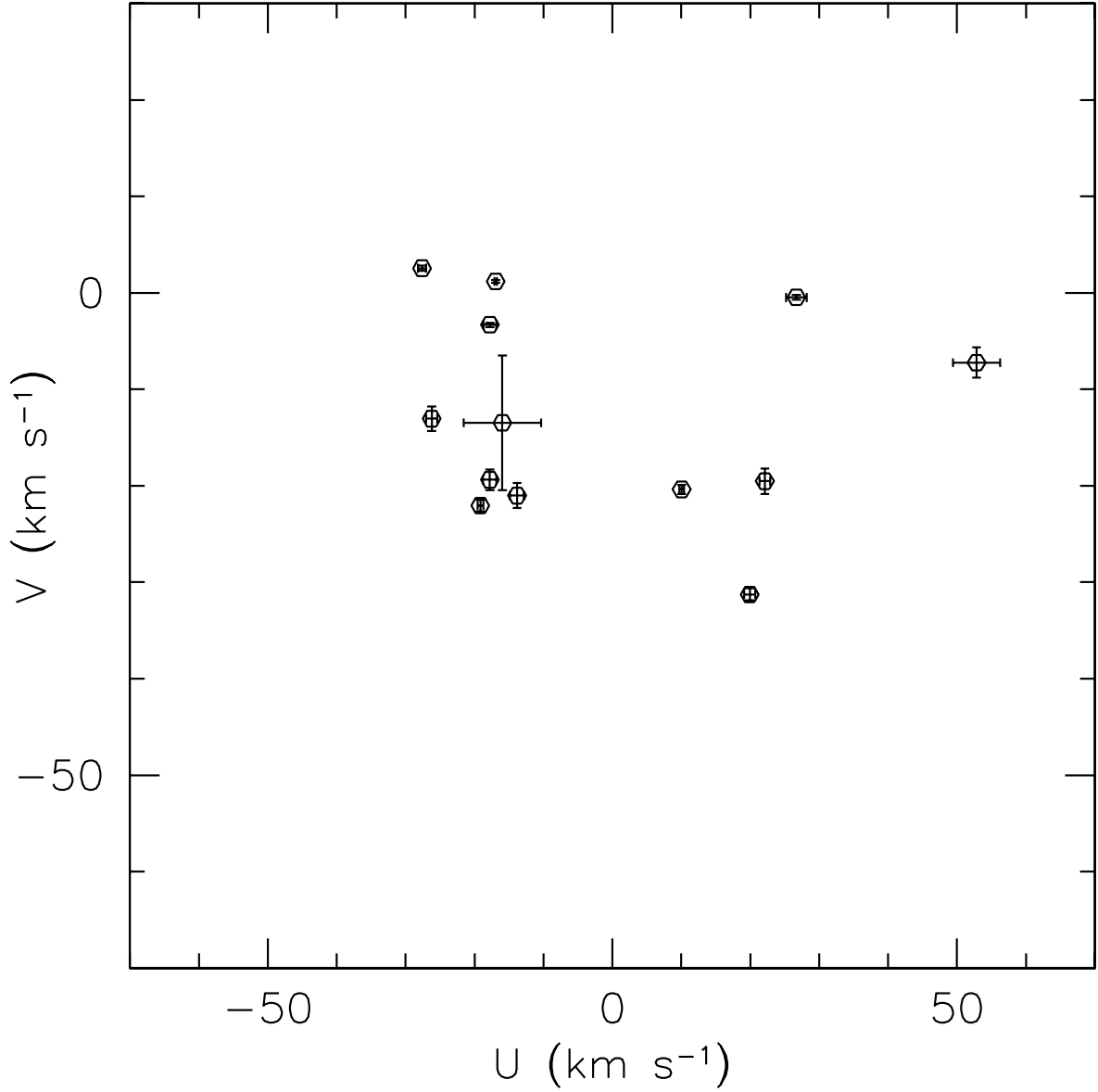}
\caption{Plot of our sample in the {\it U-V} kinematic plane.  We utilized this plot to compare with 
figures 8 and 10 from \citet{1999MNRAS.308..731S} to determine membership in moving groups of stars
in various evolutionary states.}\label{UV}
\end{figure}

\begin{figure}
\plotone{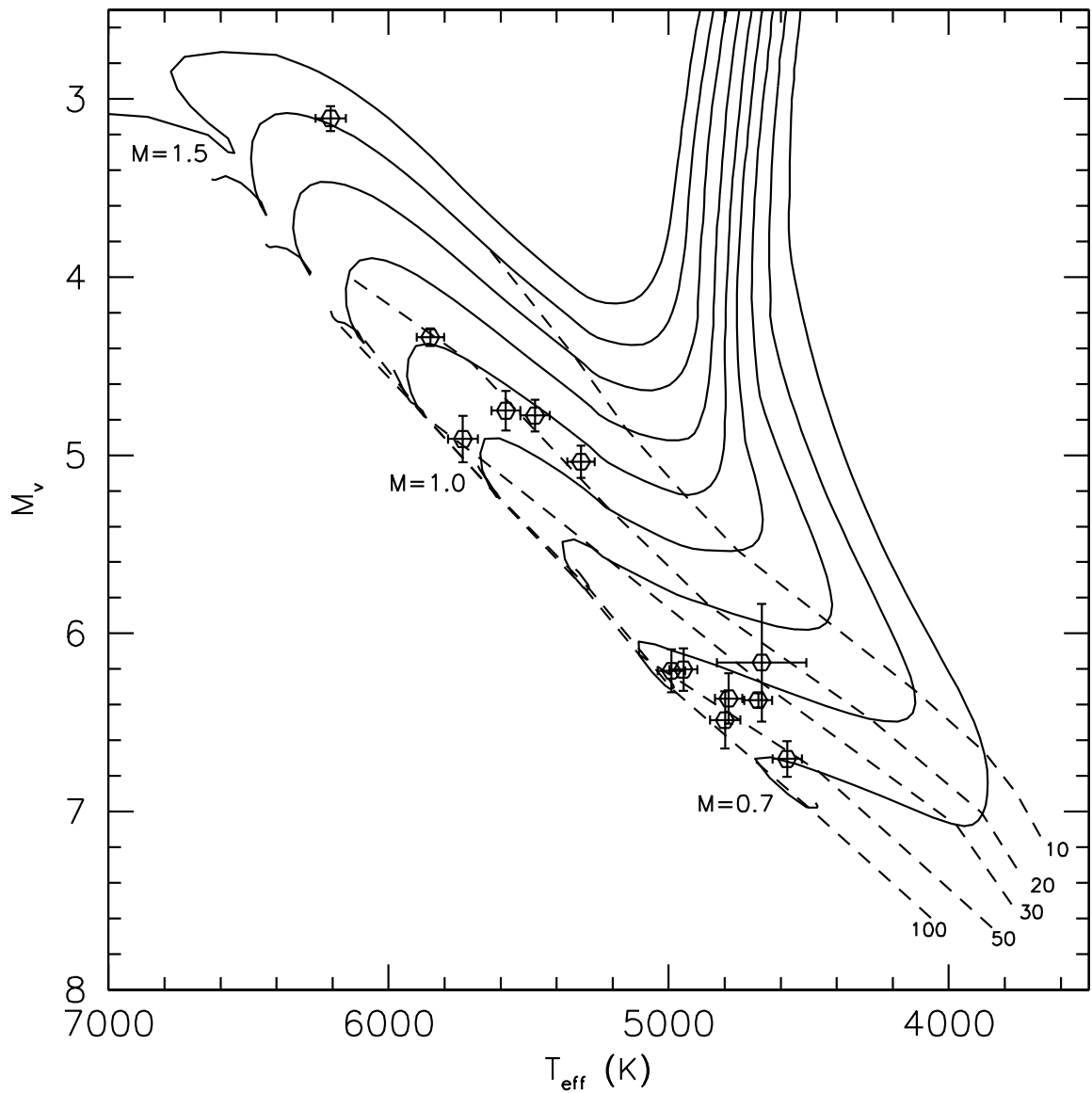}
\caption{H-R Diagram showing positions of the sample with respect to the ZAMS (100 Myr).  Mass tracks
and isochrones assuming solar metallicity are taken from \citet{1997MmSAI..68..807D}.  Ages
are estimated only for those stars which are confirmed to be post T Tauri stars
based on lithium abundances, chromospheric activity, and {\it UV} kinematics.  The isochrone ages 
are given in Myr.  The mass tracks are given in solar masses and increase in increments of 0.1
solar masses.}\label{pttcolormag}
\end{figure}

\begin{figure}
\plotone{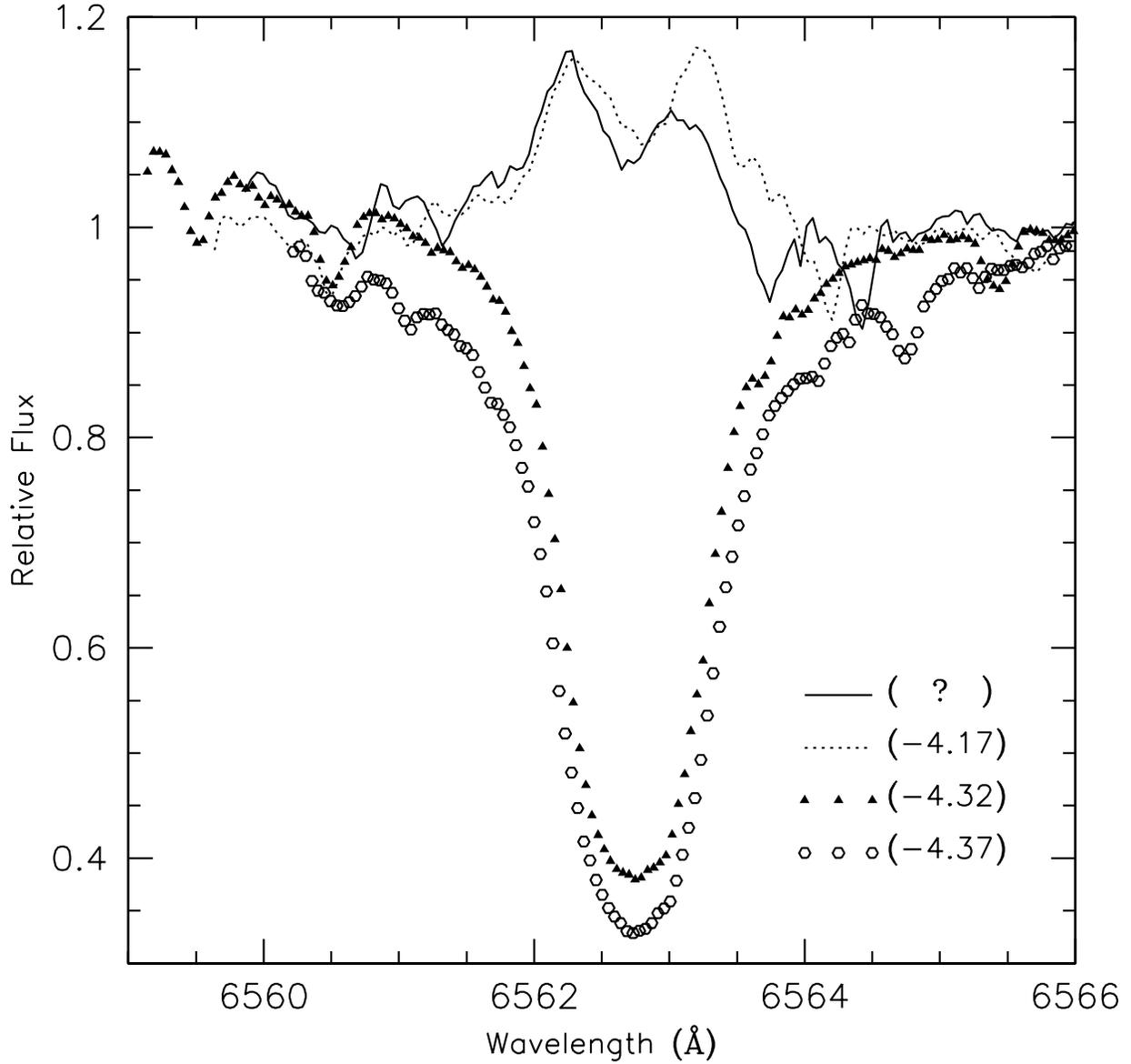}
\caption{Comparison of H$\alpha$ emission of our double lined binary with other active stars.  The legend
gives the activity levels of each line type.  The solid line is the double lined binary HIP 54529.  
The dotted line is HIP 74045, one of the candidate PTTs.  The open hexagons are HIP 87330, a ZAMS object.  
The triangles are HIP 62758, a PTT candidate with a modest level of chromospheric activity.  The 
double-lined binary clearly has H$\alpha$ in
emission, implying both youth and high chromospheric emission when compared with HIP 74045, which likewise
is in emission and has extremely high activity.}\label{halpha}
\end{figure}

\begin{figure}
\plotone{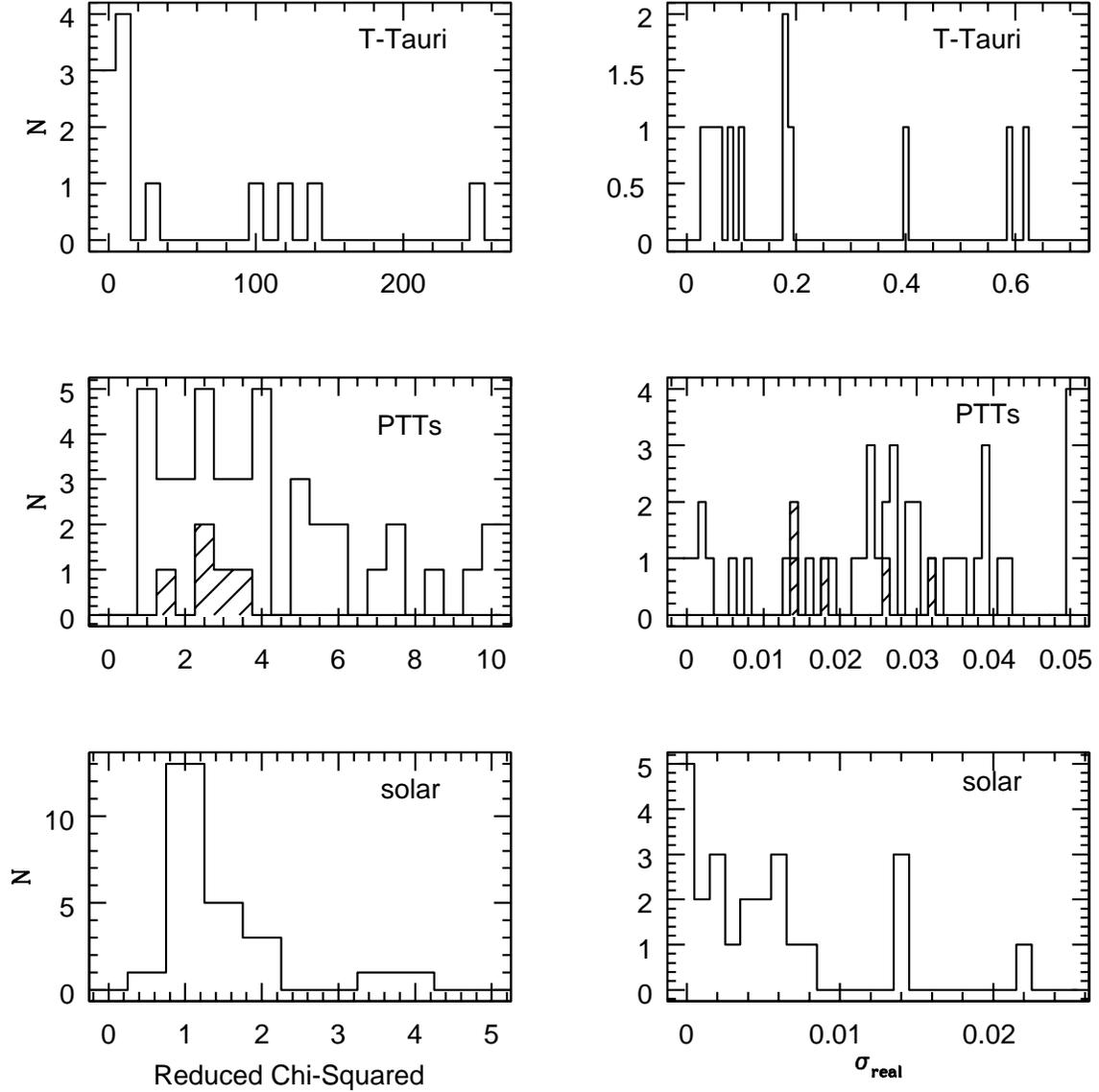}
\caption{Histograms for $\chi_{\nu}^{2}$ and real dispersions ($\sigma_{real}$) for {\it HIPPARCOS\/} V-magnitude variability.  We have
presented them in order of theoretically decreasing variability from the top down (i.e. TTs have the highest
amount of irregular variability and solar analogs the lowest).  The post T Tauri plot includes the 5 candidates 
we identify (shaded histogram) and a sample of 41 post T Tauri stars identified from the literature (unshaded histogram).}
\label{bins}
\end{figure}

\begin{figure}
\plotone{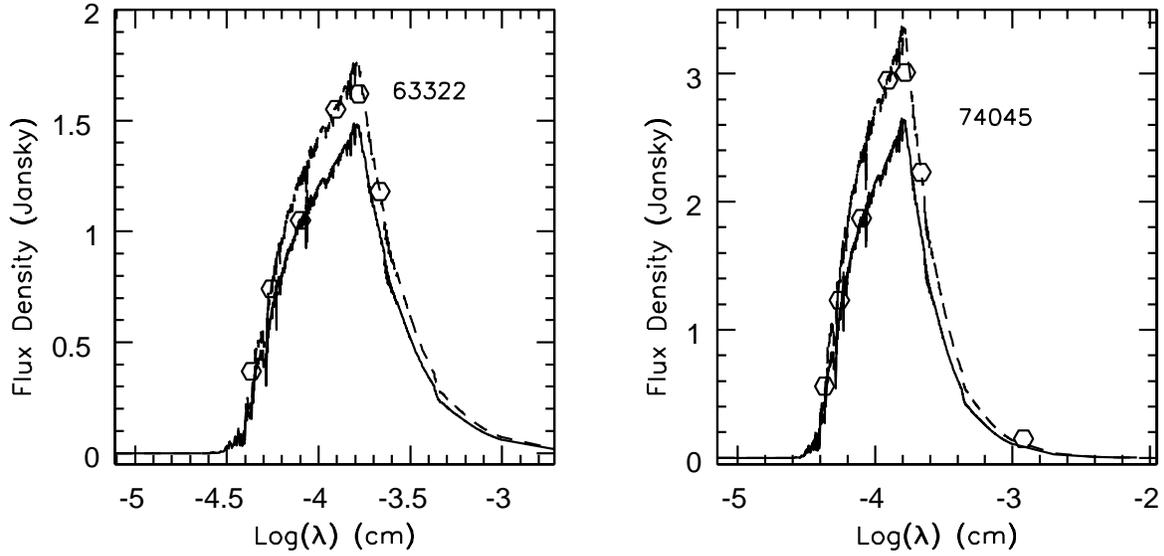}
\caption{SEDs for the two post T Tauri stars (HIP 63322 and HIP 74045) that show an infrared excess in the J, H and K bands.  
The excess appears clearly when the kurucz fluxes are normalized to the Cousins I band (solid line).  The excess, however,
is essentially non-existant when kurucz fluxes are normalized to the 2MASS J band (dotted line).  Errors are no greater 
than the size of the points.}
\label{kurucz_norm}
\end{figure}

\begin{figure}
\plotone{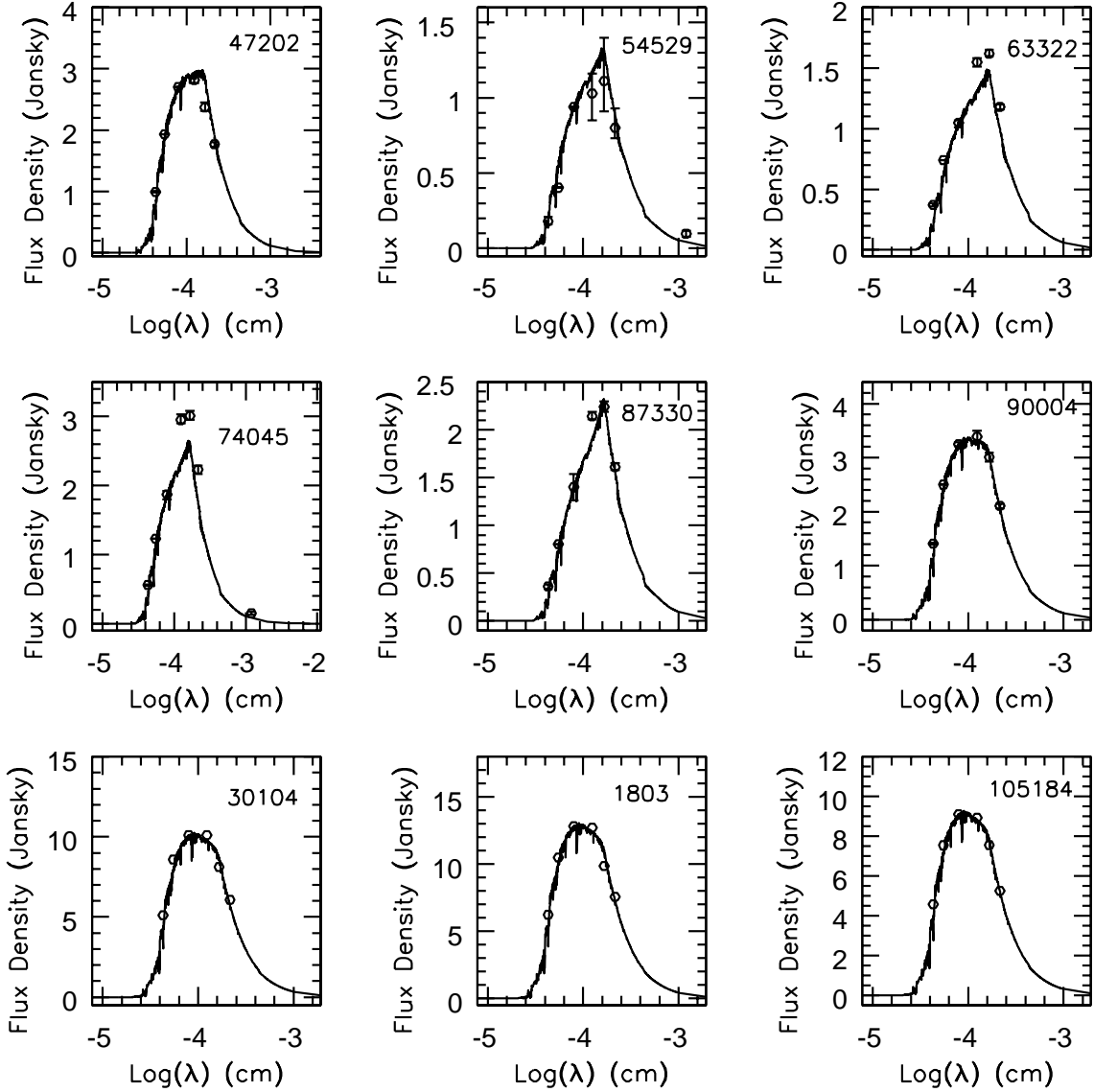}
\caption{Kurucz model atmosphere flux curves and photometric data for a selection of our stars and for a sample of solar analogs.
The top two rows are the stars in our sample and the bottom row provides curves for a sample of solar analogs.  Each plot is 
labeled with the HIP number of the corresponding star and the Kurucz fluxes are normalized to the I$_{cousins}$ band.  Note the apparent excess
in the J and H and K bands in the post T Tauri stars (HIP 63322 and HIP 745045).}
\label{blackbody1}
\end{figure}

\begin{figure}
\plotone{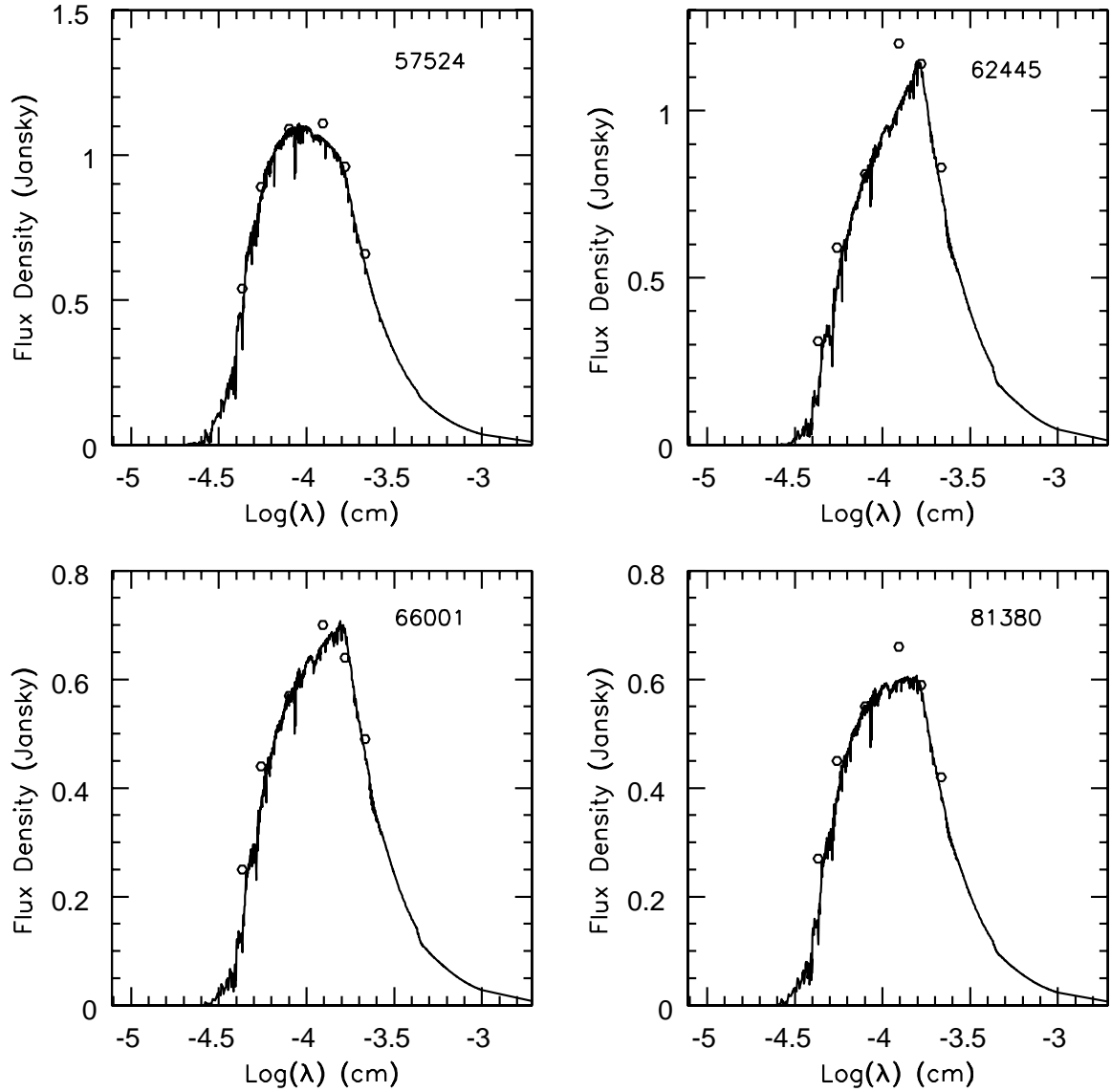}
\caption{Kurucz model atmosphere flux curves and photometric data for a sample of post T Tauri stars taken from the literature.  
HIP numbers are given inside of each of the respective plots.  We present this figure to demonstrate that post T Tauri stars
appear to demonstrate many variations of excess.  In some cases, the excesses are similar to those of our candidates (HIP 62445), although
temperature differences may suggest a different conclusion (Fig.\ref{kurucz_MAMAMIA}).}
\label{blackbody2}
\end{figure}

\begin{figure}
\plotone{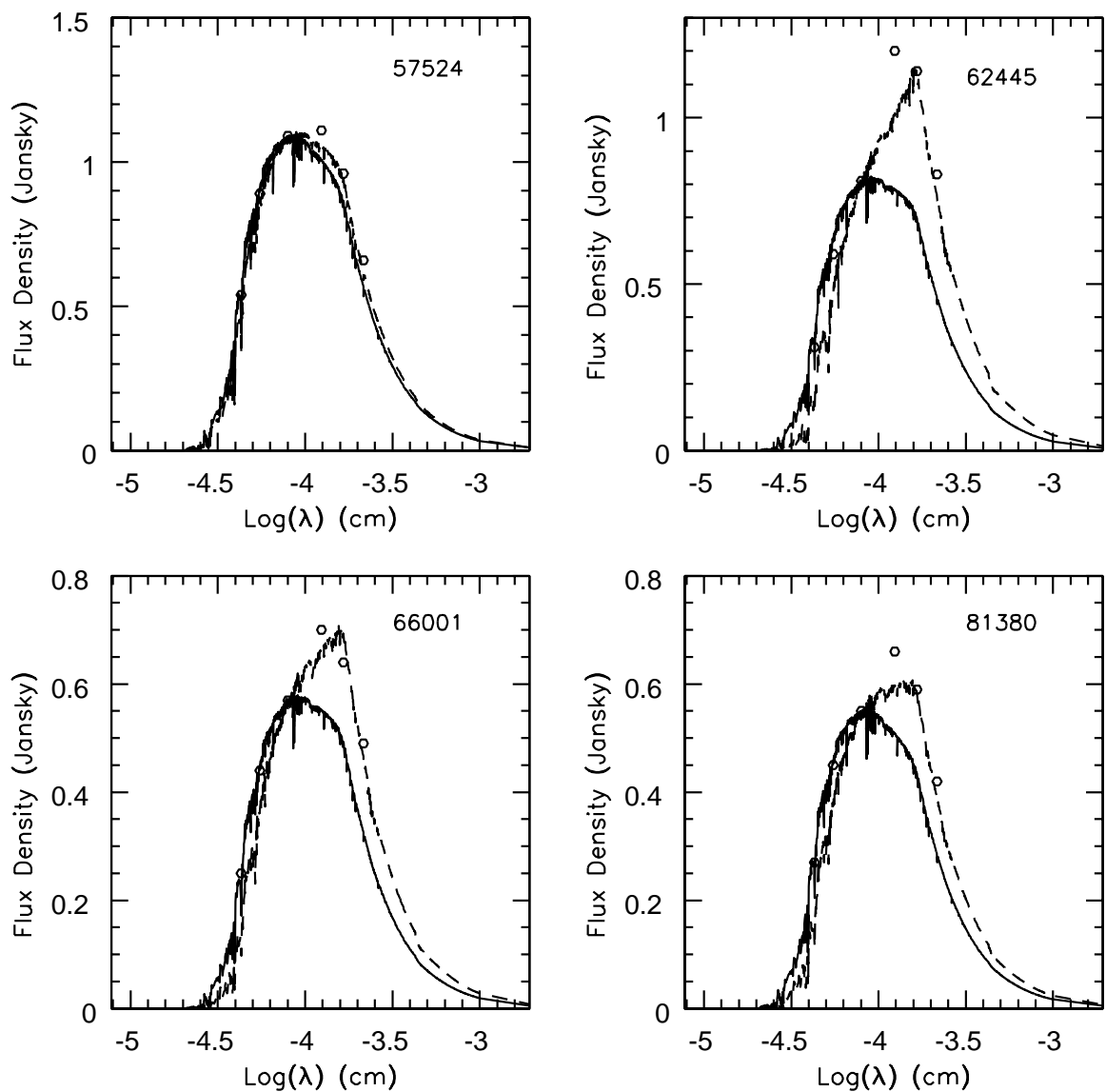}
\caption{Comparison of kurucz model flux fits to photometric data for four literature post T Tauri aged stars \citep{2002AJ....124.1670M}.
The solid curves are kurucz models for literature temperatures which were determined from calibrations to standard stars.  The 
fit appears to improve when utilizing the photometric temperatures.  We suggest that this demonstrates the usefulness of 
using photometric temperatures for the cooler stars represented by this sample.}
\label{kurucz_MAMAMIA}
\end{figure}


\begin{thebibliography}{}

\bibitem[Allende Prieto et al.(2004)]{2004A&A...420..183A} Allende Prieto, 
C., Barklem, P.~S., Lambert, D.~L., \& Cunha, K.\ 2004, \aap, 420, 183 

\bibitem[Balachandran(1995)]{1995ApJ...446..203B} Balachandran, S.\ 1995, 
\apj, 446, 203 

\bibitem[Baraffe et al.(1998)]{1998A&A...337..403B} Baraffe, I., Chabrier, 
G., Allard, F., \& Hauschildt, P.~H.\ 1998, \aap, 337, 403 

\bibitem[Boesgaard \& Tripicco(1986)]{1986ApJ...303..724B} Boesgaard, 
A.~M., \& Tripicco, M.~J.\ 1986, \apj, 303, 724 

\bibitem[Boone et al.(2006)]{2006NewAR..50..526B} Boone, R.~H., King, 
J.~R., \& Soderblom, D.~R.\ 2006, New Astronomy Review, 50, 526 

\bibitem[Butler et al.(2006)]{2006ApJ...646..505B} Butler, R.~P., et al.\ 
2006, \apj, 646, 505 

\bibitem[Carney(1983)]{1983AJ.....88..623C} Carney, B.~W.\ 1983, \aj, 88, 
623 

\bibitem[Castelli \& Kurucz(2003)]{2003IAUS..210P.A20C} Castelli, F., \& 
Kurucz, R.~L.\ 2003, Modelling of Stellar Atmospheres, 210, 20P 

\bibitem[Cayrel(1988)]{1988IAUS..132..345C} Cayrel, R.\ 1988, IAU 
Symp.~132: The Impact of Very High S/N Spectroscopy on Stellar Physics, 
132, 345 

\bibitem[Cayrel de Strobel(1996)]{1996A&ARv...7..243C} Cayrel de Strobel, 
G.\ 1996, \aapr, 7, 243 

\bibitem[Cieza et al.(2005)]{2005ApJ...635..422C} Cieza, L.~A., 
Kessler-Silacci, J.~E., Jaffe, D.~T., Harvey, P.~M., \& Evans, N.~J., II 
2005, \apj, 635, 422 

\bibitem[Cutri et al.(2003)]{2003tmc..book.....C} Cutri, R.~M., et al.\ 
2003, The IRSA 2MASS All-Sky Point Source Catalog, NASA/IPAC Infrared 
Science Archive.~http://irsa.ipac.caltech.edu/applications/Gator/,

\bibitem[D'Antona \& Mazzitelli(1997)]{1997MmSAI..68..807D} D'Antona, F., 
\& Mazzitelli, I.\ 1997, Memorie della Societa Astronomica Italiana, 68, 
807 

\bibitem[Eggen(1969)]{1969PASP...81..553E} Eggen, O.~J.\ 1969, \pasp, 81, 
553 

\bibitem[Fitzpatrick \& Sneden(1987)]{1987BAAS...19.1129F} Fitzpatrick, 
M.~J., \& Sneden, C.\ 1987, \baas, 19, 1129 

\bibitem[Gray et al.(2003)]{2003AJ....126.2048G} Gray, R.~O., Corbally, 
C.~J., Garrison, R.~F., McFadden, M.~T., \& Robinson, P.~E.\ 2003, \aj, 
126, 2048

\bibitem[Henry et al.(1996)]{1996AJ....111..439H} Henry, T.~J., Soderblom, 
D.~R., Donahue, R.~A., \& Baliunas, S.~L.\ 1996, \aj, 111, 439 

\bibitem[Herbig(1978)]{1978ppeu.book..171H} Herbig, G.~H.\ 1978, Problems 
of Physics and Evolution of the Universe, 171 

\bibitem[Herbig \& Bell(1995)]{1995yCat.5073....0H} Herbig, G.~H., \& Bell, 
K.~R.\ 1995, VizieR Online Data Catalog, 5073, 0 

\bibitem[Israelian et al.(2004)]{2004A&A...414..601I} Israelian, G., 
Santos, N.~C., Mayor, M., \& Rebolo, R.\ 2004, \aap, 414, 601 

\bibitem[Jensen(2001)]{2001ASPC..244....3J} Jensen, E.~L.~N.\ 2001, ASP 
Conf.~Ser.~244: Young Stars Near Earth: Progress and Prospects, 244, 3 

\bibitem[Johnson \& Soderblom(1987)]{1987AJ.....93..864J} Johnson, 
D.~R.~H., \& Soderblom, D.~R.\ 1987, \aj, 93, 864 

\bibitem[Jones et al.(1996)]{1996AJ....112..186J} Jones, B.~F., Shetrone, 
M., Fischer, D., \& Soderblom, D.~R.\ 1996, \aj, 112, 186 

\bibitem[King et al.(1997)]{1997AJ....113.1871K} King, J.~R., Deliyannis, 
C.~P., Hiltgen, D.~D., Stephens, A., Cunha, K., \& Boesgaard, A.~M.\ 1997, 
\aj, 113, 1871

\bibitem[King et al.(2000)]{2000AJ....119..859K} King, J.~R., 
Krishnamurthi, A., \& Pinsonneault, M.~H.\ 2000, \aj, 119, 859 

\bibitem[Mamajek et al.(2002)]{2002AJ....124.1670M} Mamajek, E.~E., Meyer, 
M.~R., \& Liebert, J.\ 2002, \aj, 124, 1670 

\bibitem[Nordstr{\"o}m et al.(2004)]{2004A&A...418..989N} Nordstr{\"o}m, 
B., et al.\ 2004, \aap, 418, 989 

\bibitem[Perryman \& ESA(1997)]{1997hity.book.....P} Perryman, M.~A.~C., \& 
ESA 1997, The Hipparcos and Tycho catalogues.~Astrometric and photometric 
star catalogues derived from the ESA {\it HIPPARCOS\/} Space Astrometry Mission, 
Publisher: Noordwijk, Netherlands: ESA Publications Division, 1997, Series: 
ESA SP Series vol no: 1200, ISBN: 9290923997 (set),

\bibitem[Petrov(2003)]{2003Ap.....46..506P} Petrov, P.~P.\ 2003, 
Astrophysics, 46, 506 

\bibitem[Ram{\'{\i}}rez \& Mel{\'e}ndez(2005)]{2005ApJ...626..465R} 
Ram{\'{\i}}rez, I., \& Mel{\'e}ndez, J.\ 2005, \apj, 626, 465 

\bibitem[Saffe et al.(2005)]{2005A&A...443..609S} Saffe, C., G{\'o}mez, M., 
\& Chavero, C.\ 2005, \aap, 443, 609 

\bibitem[Santos et al.(2004)]{2004A&A...415.1153S} Santos, N.~C., 
Israelian, G., \& Mayor, M.\ 2004, \aap, 415, 1153 

\bibitem[Schuler et al.(2006)]{2006ApJ...636..432S} Schuler, S.~C., King, 
J.~R., Terndrup, D.~M., Pinsonneault, M.~H., Murray, N., \& Hobbs, L.~M.\ 
2006, \apj, 636, 432 

\bibitem[Siess et al.(2000)]{2000A&A...358..593S} Siess, L., Dufour, E., \& 
Forestini, M.\ 2000, \aap, 358, 593 

\bibitem[Skuljan et al.(1999)]{1999MNRAS.308..731S} Skuljan, J., Hearnshaw, 
J.~B., \& Cottrell, P.~L.\ 1999, \mnras, 308, 731 

\bibitem[Sneden(1973)]{1973ApJ...184..839S} Sneden, C.\ 1973, \apj, 184, 
839 

\bibitem[Soderblom et al.(1993)]{1993AJ....106.1059S} Soderblom, D.~R., 
Jones, B.~F., Balachandran, S., Stauffer, J.~R., Duncan, D.~K., Fedele, 
S.~B., \& Hudon, J.~D.\ 1993, \aj, 106, 1059

\bibitem[Soderblom et al.(1998)]{1998ApJ...498..385S} Soderblom, D.~R., et 
al.\ 1998, \apj, 498, 385 

\bibitem[Zuckerman \& Song(2004)]{2004ARA&A..42..685Z} Zuckerman, B., \& 
Song, I.\ 2004, \araa, 42, 685 

\end{thebibliography}
\end{document}